\documentclass[runningheads, envcountsame, a4paper]{llncs}

\usepackage[pdftex]{graphicx}
\usepackage{balance}  
\usepackage{cite}

\usepackage{xcolor}
\usepackage{color, colortbl}
\usepackage{algorithmic}
\usepackage{amsmath}
\usepackage[hyphens]{url}
\usepackage{multirow}
\usepackage[boxruled, linesnumbered] {algorithm2e}
\usepackage{parskip}
\usepackage{mathtools}
\usepackage{amsfonts}
\usepackage{threeparttable}
\usepackage[normalem]{ulem}
\usepackage{booktabs}
\usepackage{epstopdf}

\newtheorem{Definition}{Definition}[section]
\newcommand{\nop}[1]{}

\newcommand{\system}{{\sc VeriDL}\xspace}
\def\BibTeX{{\rm B\kern-.05em{\sc i\kern-.025em b}\kern-.08em
    T\kern-.1667em\lower.7ex\hbox{E}\kern-.125emX}}

\newcommand{\Boxiang}[1]{{{\textcolor{black}{\textbf{Boxiang:}}}{\textcolor{blue}{\textbf{#1}}}}}
\newcommand{\Wendy}[1]{{{\textcolor{black}{\textbf{Wendy:}}}{\textcolor{red}{\textbf{#1}}}}}
\newcommand{\Bo}[1]{{{\textcolor{black}{\textbf{Bo:}}}{\textcolor{green}{\textbf{#1}}}}}

\newenvironment{ditemize}{%
\begin{list}{{\bf $\bullet$}}{%
\setlength{\itemsep}{0pt}\setlength{\rightmargin}{0pt}%
\setlength{\leftmargin}{1.2em}\setlength{\parsep}{0pt}}}{
\end{list}}

\begin{document}

\title{VeriDL: Integrity Verification of Outsourced Deep Learning Services (Extended Version)}

\author{Anonymous}
\institute{}

\author{Boxiang Dong\inst{1} \and
Bo Zhang\inst{2} \and
Hui (Wendy) Wang\inst{3}}
\authorrunning{B. Dong, and B. Zhang, and H. Wang}
%
\institute{Montclair State University, Montclair, NJ  \email{dongb@montclair.edu} \and
Amazon Inc., Seattle, WA \email{bzhanga@amazon.com} \and
Stevens Institute of Technology, Hoboken, NJ \email{Hui.Wang@stevens.edu}
}

\maketitle

\setcounter{footnote}{0}

\begin{abstract} 
Deep neural networks (DNNs) are prominent due to their superior performance in many fields. 
The deep-learning-as-a-service (DLaaS) paradigm  enables individuals and organizations (clients) to outsource their DNN learning tasks to the cloud-based platforms. 
However, the DLaaS server may return incorrect  DNN models due to various reasons (e.g., Byzantine failures). 
This raises the serious concern of how to verify if the DNN models trained by potentially untrusted DLaaS servers are indeed correct.
To address this concern, in this paper, we design \system, a framework that supports efficient correctness verification of DNN models in the DLaaS paradigm. 
The key idea of \system is the design of a small-size cryptographic proof of the training process of the DNN model, which is  associated with the model and returned to the client. Through the proof, \system can verify the correctness of the DNN model returned by the DLaaS server with a deterministic guarantee and cheap overhead. 
Our experiments on four real-world datasets  demonstrate the efficiency and effectiveness of \system.
\vspace{-0.1in}
\keywords{Deep learning, integrity verification, deep-learning-as-a-service}
\end{abstract}

\vspace{-0.1in}
\section{Introduction}
The recent abrupt advances in deep learning (DL) \cite{bengio2009learning,lecun2015deep} have 
led to breakthroughs in many  fields such as speech recognition, image classification, text translation, etc.  
However, this success crucially relies on the availability of both hardware and software resources, as well as human expertise for many learning tasks. As the complexity of these tasks is often beyond non-DL-experts, the rapid growth of DL  applications has created a demand for cost-efficient, off-shelf solutions. This motivated the emerge of the {\em deep-learning-as-a-service} (DLaaS) paradigm which enables individuals and organizations (clients) to outsource their data and deep learning tasks to the cloud-based service providers for their needs of  flexibility, ease-of-use, and cost efficiency. 


Despite providing cost-efficient DL solutions, outsourcing training to DLaaS service providers raises serious security and privacy concerns. One of the major issues is the {\em integrity}  of the deep neural network (DNN) models trained by the server. 
For example,  due to Byzantine failures such as software bugs and network connection interruptions, the server may return a DNN model that does not reach its convergence. However, it is difficult for the client to verify the correctness of the returned DNN model easily due to the lack of hardware resources and/or DL expertise. 

In this paper, we consider the {\em Infrastructure-as-a-Service} (IaaS) setting  where the DLaaS service provider delivers the  computing infrastructure including servers, network, operating systems, and storage as the services through virtualization technology. Typical examples of IaaS settings are Amazon Web Services\footnote{Amazon Web Services: https://aws.amazon.com/} and Microsoft Azure\footnote{Microsoft Azure: https://azure.microsoft.com/en-us/}.
In this setting, a client outsources his/her training data $T$ to the DLaaS service provider (server). The client does not need to store $T$ locally after it is outsourced (i.e., the client may not have access to $T$ after outsourcing). The client also has the complete control of the infrastructure. He can customize the configuration of the DNN model $M$, including the network topology and hyperparameters of $M$. Then the server trains $M$ on the outsourced $T$ and returns the trained model $M$ to the client. 
As the client lacks hardware resources and/or DL expertise, a third-party verifier will authenticate on behalf of the client if $M$ returned by the server is {\em correct}, i.e., $M$ is the same as being trained locally with $T$ under the same configuration. 
Since the verifier may not be able to access the private training data owned by the client, our goal is to design a lightweight verification mechanism that enables the verifier to authenticate the correctness of $M$ without full access to $T$. 
A possible solution is that the {\em verifier} executes the training process independently. Since the verifier does not have the access to the client's private data, he has to execute training on the private data encrypted by homomorphic encryption (HE) \cite{gilad2016cryptonets,hesamifard2017cryptodl}. Though correct, this solution can incur expensive overhead due to the high complexity of HE. 
Furthermore,  since HE only supports polynomial functions, 
 some activation functions  (e.g., ReLU, Sigmoid, and Tanh) have to be approximated by low-degree polynomials when HE is used, and thus the verifier cannot compute the exact model updates. 
On the other hand,  the existing works on verifying the integrity of DNNs (e.g.,  SafetyNets \cite{ghodsi2017safetynets} and VeriDeep \cite{he2018verideep} hold a few restrictions on the activation function (e.g., it must be polynomials with integer coefficients) and data type of weights/inputs (e.g., they must be integers). We do not have any assumption on activation functions and input data types. Furthermore, these existing works have to access the original data, which is prohibited in our setting due to privacy protection.

{\bf Our contributions.} 
We design \system, a framework that supports efficient verification of outsourced DNN model training by a potentially untrusted DLaaS server which may  return wrong DNN model as the result. 
\system provides the {\em deterministic}  correctness guarantee of remotely trained DNN models without any constraint on the activation function and the types of input data. 
The key idea of \system is that the server constructs a cryptographic proof of the model updates, 
and sends the proof along with the model updates to the verifier. 
Since the proof aggregates the intermediate model updates (in compressed format) during training, the verifier can authenticate the correctness of the trained model by using the proof only. 
In particular, we make the following contributions.  
\underline{First}, we design an efficient procedure to  construct the cryptographic proof whose size  is significantly smaller than the training data. The proof is constructed by using {\em bilinear pairing}, which is a cryptographic protocol commonly used for aggregate signatures.
\underline{Second}, we design a lightweight verification method named \system that can authenticate the correctness of model updates through the  cryptographic proof. By using the proof, \system does not need  access to the training data for correctness verification. 
\underline{Third}, 
as the existing bilinear mapping methods cannot deal with the weights in DNNs that are decimal or negative values, we significantly extend the bilinear mapping protocol to handle decimal and negative values. 
We formally prove that  \system is secure against the attacker who may have full knowledge of the verification methods and thus try to escape from verification. 
\underline{Last but not least}, we implement the prototype of \system, deploy it on a DL system, and evaluate its performance on four real-world datasets that are of different data types (including non-structured images and structured tabular data). Our experimental results demonstrate the efficiency and effectiveness of \system. The verification by \system is faster than the existing DNN verification methods   \cite{gilad2016cryptonets,hesamifard2017cryptodl} by more than three orders of magnitude.

\section{Preliminaries}
\label{sc:pre}

 {\bf Bilinear mapping.} Let $G$ and $G_T$ be two multiplicative cyclic groups of finite order $p$.  Let $g$ be a generator of $G$. 
 A bilinear group mapping $e$ is defined as $e: G \times G \rightarrow G_T$, which has the following property: $\forall a, b \in \mathbb{Z}_p$, $e(g^a , g^b) = e(g, g)^{ab}$. In the following discussions, we use the terms bilinear group mapping and bilinear mapping interchangeably. 
 The main advantage of bilinear mapping is that determining whether $c \equiv ab\ \textsf{mod}\ p$ 
 {\em without the access to} $a$, $b$ and $c$ can be achieved by checking whether $e(g^a, g^b) = e(g, g^c)$, by 
 given $g, g^a, g^b, g^c$. 
 
\label{sc:framework}
{\bf Outsourcing framework.} We consider the outsourcing paradigm that involves three parties: (1) a {\em data owner} (client) $\mathcal{O}$ who holds a private training dataset $T$; 
 (2) a third-party service provider (server) $\mathcal{S}$ who provides infrastructure services to $\mathcal{O}$; 
 and (3) 
 a third-party verifier $\mathcal{V}$ who authenticates the integrity of $\mathcal{S}$' services. In this paradigm, $\mathcal{O}$ outsources $T$ to $\mathcal{S}$ for training of a DNN model $M$. Meanwhile $\mathcal{O}$ specifies the configuration of $M$ on $\mathcal{S}$' infrastructure for training of $M$. After $\mathcal{S}$ finishes training of $M$, it sends $M$ to $\mathcal{V}$ for verification. Due to privacy concerns, $\mathcal{V}$ cannot access  the private training data $T$ for verification. 
{\bf Basic DNN operations.}
In this paper, we only focus on deep feedforward networks (DNNs), and leave more complicated structures like convolutional and recurrent networks for the future work.
In this section, we present the basic operations of training a DNN model. We will explain in Section \ref{sc:deterministic} how to verify the output of these operations. 
In this paper, we only concentrate on fully-connected neural networks, and refrain from convolutional networks or recurrent networks. However, our design can be adapted to more advanced network structures.

A DNN consists of several layers, including the input layer (data samples), the output layer (the predicted labels), and a number of hidden layers.
During the feedforward computation, for the neuron $n_k^{\ell}$, its weighted sum $z^{\ell}_k$ is defined as:
\begin{equation}
    \label{eq:z}
    z^{\ell}_k = 
    \begin{cases}
        \sum_{i=1}^{m} x_i w_{ik}^{\ell} & \text{if } {\ell}=1 \\
        \sum_{j=1}^{d_{\ell-1}} a_j^{\ell-1} w_{jk}^{\ell} & \text{otherwise,}
    \end{cases}
\end{equation}
where $x_i$ is the $i$-th feature of the input $\vec{x}$, and $d_{i}$ is the number of neurons on the $i$-th hidden layer. The activation $a^{\ell}_k$ is calculated as follows:
\begin{equation}
    \label{eq:a}
    a^{\ell}_k = \sigma(z^{\ell}_k),
\end{equation}
where $\sigma$ is the activation function. We allow a broad class of activation functions such as sigmoid, ReLU (rectified linear unit), and hyperbolic tangent. 

On the output layer, the output $o$ is generated by following: 
\begin{equation}
   o = \sigma(z^o) = \sigma(\sum_{j=1}^{d_{L}} a^{L}_j w_{j}^o),
    \label{eq:forward_output}
    \vspace{-0.1in}
\end{equation}
where $w^o_{j}$ is the weight that connects $n^{\ell}_j$ to the output neuron.

In this paper, we mainly consider the mean square error (MSE) as the cost function. 
For any sample $(\vec{x}, y)\in T$, the cost $C(\vec{x}, y; W)$ is measured as the difference between the label $y$ and the output $o$: 
\begin{equation}
    \begin{split}
    C(\vec{x}, y; W) & = C(o, y) 
    = \frac{1}{2} (y- o)^2.
    \end{split}
    \label{eq:cost}
\end{equation}
Then the  error $E$ is calculated as the average error for all samples:  
\begin{equation}
    E = \frac{1}{N} \sum_{(\vec{x}, y) \in T} C(\vec{x}, y; W).
\label{eq:error}
\end{equation}

In the backpropagation process, gradients are calculated to update the weights in the neural network. According to the chain rule of backpropagation \cite{lecun2015deep}, for any sample $(\vec{x}, y)$, the error signal $\delta^o$ on the output neuron is 
\begin{equation}
    \delta^o = \nabla_{o} C(o, y) \odot \sigma'(z^o) = (o - y) \sigma'(z^o).
    \label{eq:signal_1}
\end{equation}
While the error signal $\delta_k^{\ell}$ at the $\ell$-th hidden layer is 
\begin{equation}
    \delta_k^{\ell} =
    \begin{cases}
    \sigma'(z^{\ell}_k) w^{o}_{k} \delta^{o} & \text{if } \ell=L, \\
    \sigma'(z^{\ell}_k) \sum_{j=1}^{d_{\ell+1}} w^{\ell+1}_{kj} \delta^{\ell+1}_j & \text{otherwise}. 
    \end{cases}
    \label{eq:signal_2}
\end{equation}
where $\ell=L$ indicates the last hidden layer.

The derivative for each weight $w_{jk}^{\ell}$ is computed as: 
\begin{equation}
    \frac{\partial C}{\partial w_{jk}^{\ell}} = 
    \begin{cases}
    x_j \delta_{k}^{\ell} & \text{if } \ell=1 \\
    a_j^{\ell-1} \delta_{k}^{\ell} & \text{otherwise.} 
    \end{cases}
    \label{eq:gradient}
    \vspace{-0.1in}
\end{equation}

Then the weight increment $\Delta w_{jk}^{\ell}$ is 
\begin{equation}
    \Delta w_{jk}^{\ell} = -\frac{\eta}{N} \sum_{(\vec{x}, y) \in T} \frac{\partial C}{\partial w_{jk}^{\ell}},
    \label{eq:increment}
\end{equation}
where $\eta$ is the learning rate. Finally, the weight is updated as 
\begin{equation}
    w_{jk}^{\ell} = w_{jk}^{\ell} + \Delta w_{jk}^{\ell}.
    \label{eq:update}
\end{equation}

The DNN is iteratively optimized by following the above feedforward and backpropagation process until it reaches convergence,  $|E_1-E_2|\leq \theta$, where $E_1$ and $E_2$ are the error/loss of two consecutive epochs in the optimization process, and $\theta$ is a small constant.

{\bf Verification protocol. }
We adapt the definition of the integrity verification protocol  \cite{papamanthou2011optimal} to our setting: \begin{Definition}[Deep Learning Verification Protocol]
\label{def:protocol}
Let $W$ be the set of weight parameters in a DNN, and $T$ be a collection of data samples. Let $\Delta W$ be the parameter update after training the DNN on $T$. The authentication protocol is a collection of the following four polynomial-time algorithms: $\mathbf{genkey}$ for key generation, $\mathbf{setup}$ for initial setup, $\mathbf{certify}$ for verification preparation, and $\mathbf{verify}$ for verification.
\begin{itemize}
\item \{$s_k, p_k$\} $\leftarrow \mathbf{genkey(})$: It outputs a pair of secret and public key;
\item \{$\gamma$\} $\leftarrow \mathbf{setup(T, s_k, p_k})$: Given the dataset $T$, the secret key $s_k$ and the public key $p_k$, it returns a single signature $\gamma$ of $T$; 
\item \{$\pi$\} $\leftarrow \mathbf{certify(T, W_0, \Delta W, p_k)}$:  Given the data collection $T$, the initial DNN model parameters $W_0$, the model update $\Delta W$, and a public key $p_k$, it returns the proof $\pi$; 
\item \{$\mathsf{accept, reject}$\} $\leftarrow$ $\mathbf{verify(W_0, \Delta W, \pi, \gamma, p_k)}$:  Given the initial DNN model parameters $W_0$, the  model update $\Delta W$, the proof $\pi$, the signature $\gamma$, and the public key $p_k$, it outputs either $\mathsf{accept}$ or $\mathsf{reject}$. 
\end{itemize}
\end{Definition}

In this paper, we consider the adversary who has full knowledge of the authentication protocol. 
Next, we define the security of the authentication protocol against such adversary. 

\begin{Definition}
\label{def:security}
Let $\mathbf{Auth}$ be an authentication scheme \{$\mathbf{genkey}, \mathbf{setup}, \mathbf{certify},$ $\mathbf{verify}$\}. Let $\mathbf{Adv}$ be a probabilistic polynomial-time adversary that is only given $p_k$ and has unlimited access to all
algorithms of $\mathbf{Auth}$. Then, given a DNN with initial parameters $W_0$ and a dataset $T$, $\mathbf{Adv}$ returns a wrong model update $\Delta W'$ and a proof $\pi'$:  
$\{\Delta W', \pi'\} \leftarrow \mathbf{Adv}(D, W_0, p_k)$. 
We say $\mathbf{Auth}$ is secure if for any $p_k$ generated by the $\mathbf{genkey}$ routine, for any $\gamma$ generated by the $\mathbf{setup}$ routine, and for any probabilistic polynomial-time adversary 
$\mathbf{Adv}$, it holds that 
\[Pr (accept\leftarrow\mathbf{verify}(W_0, \Delta W', \pi', \gamma, p_k)) \leq  negli(\lambda),\]
\end{Definition}
where $negli(\lambda)$ is a negligible function in the security parameter $\lambda$. 
Intuitively, $\mathbf{Auth}$ is secure if with negligible probability the incorrect model update can  escape from verification. 

\section{Problem Statement}
\label{sc:ps}

{\bf Threat model.} In this paper, we consider the server $\mathcal{S}$ that may return incorrect trained model due to various reasons. 
For example, the learning process might be terminated before it reaches convergence due to the system's Byzantine failures (e.g., software bugs and network issues). 
$\mathcal{S}$ may also be incentivized to halt the training program early in order to save the computational cost and seek for a higher profit.
Given the untrusted nature of the remote server, it is thus crucial for the client to verify the correctness of the returned DNN model before using the model for any decision-making task. 

{\bf Problem statement.} We consider the problem setting in which the data owner $\mathcal{O}$ outsources the  training set $T$ on the server. $\mathcal{O}$  also can specify the configuration of the DNN model $M$ whose initial parameters are specified by $W_0$.
The server $\mathcal{S}$ trains $M$ until it reaches convergence (a local optima), and outputs the model update $\Delta W = f(T; W_0)$. 
However, with the presence of security threats, the model update $\Delta W$ returned by the server may not be a local optima. Therefore, our goal is to design an integrity verification protocol (Def. \ref{def:protocol}) that enables a third-party verifier $\mathcal{V}$ to verify if $\Delta W$ helps the model reach convergence without the access to the private training data. 
\section{Authentication Method}
\label{sc:deterministic}
In this section, we explain the details of our authentication protocol. 
The $\bf{genkey}$ protocol is straightforward: the data owner $\mathcal{O}$ picks a pairing function $e$ on two sufficiently large cyclic groups $G$ and $G_T$ of order $p$, a generator $g\in G$, and a secret key $s\in \mathbb{Z}_p$. Then it outputs a pair of secrete and public key $(s_k, p_k)$,  where $s_k=s$, and $p_k=\{g, G, G_T, e, v, H(\cdot)\}$, 
where $v=g^s \in G$, and $H(\cdot)$ is a hash function whose output domain is $\mathbb{Z}_p$.  $\mathcal{O}$ keeps $s_k$ private and distributes $p_k$ to the other involved parties. 
In the following discussions, we only focus on the $\mathbf{setup}$, $\mathbf{certify}$ and $\mathbf{verify}$ protocols. 

\nop{
\begin{figure}[t!]
\begin{center}
\includegraphics[width=0.85\textwidth]{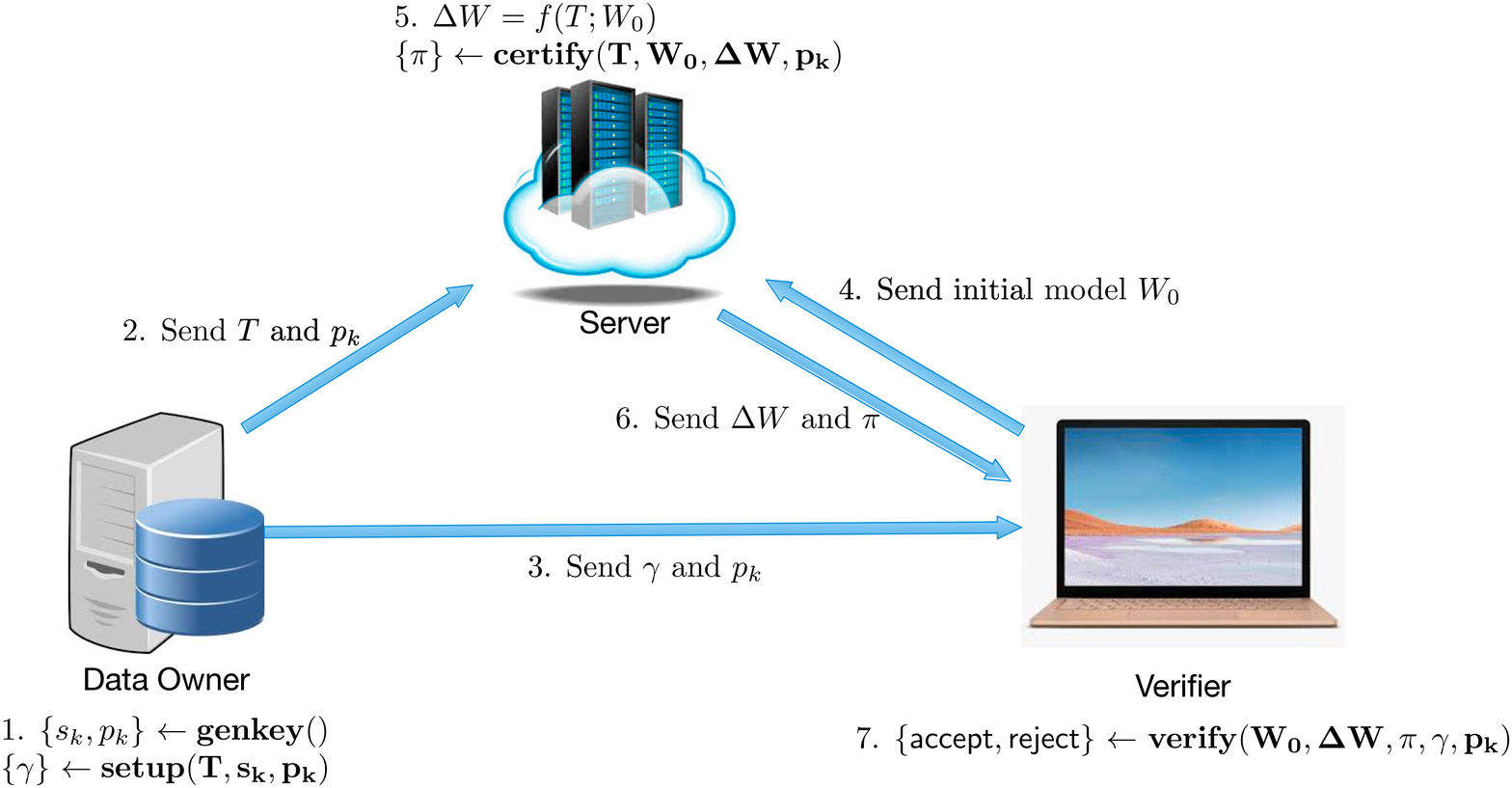}
\caption{\label{fig:frameworkD} Overview of \system \Wendy{1. Change "verifier" to "Verifier". 2. Change all "D" to "T"}} \Boxiang{Fixed} 
\end{center}
\end{figure}
}

{\bf Overview of our Approach.} 
 We design a verification method that only uses a short proof of the results for verification. 
Consider a data owner $\mathcal{O}$ that has a private  dataset $T$. Before transferring $T$ to the server,  $\mathcal{O}$  executes the {\em setup} protocol to generate a short signature $\gamma$ of $T$, and disseminate $\gamma$ to the verifier $\mathcal{V}$. 
$\mathcal{O}$ also sets up a DNN model $M$ with initial weights $W_0$. Then $\mathcal{O}$ outsources $M$ (with $W_0)$ and the training dataset $T$ to $\mathcal{S}$. 
After receiving $T$ and $M$ with its initial setup, the server $\mathcal{S}$  optimizes $M$  and obtains the model updates $\Delta W$. Besides returning $\Delta W$ to the verifier $\mathcal{V}$, $\mathcal{S}$  sends two errors $E_1$ and $E_2$, where $E_1$ is the error when the model reaches convergence as claimed (computed by Eqn. \ref{eq:error}) and $E_2$ is the error by running an additional round of backpropagation and feedforward process after convergence. Furthermore, $\mathcal{S}$  follows the {\em certify} protocol and constructs a short cryptographic proof $\pi$ of $E_1$ and $E_2$. 
The proof $\pi$ includes: (1) the cryptographic digest $\pi_T$ of the samples, and (2) the  intermediate results of feedforward and backpropagation processes in computing $E_1$ and $E_2$. The verifier $\mathcal{V}$  then runs the {\em verify} protocol and checks the correctness of $\Delta W$ by the following three steps:  
\begin{ditemize}
    \item {\em Authenticity verification of $\pi_T$}: $\mathcal{V}$ checks the integrity of $\pi_T$ against the dataset signature $\gamma$ that is signed by $\mathcal{O}$;
    \item {\em Authenticity verification of $E_1$ and $E_2$}: Without access to the private data $T$, $\mathcal{V}$ verifies if both errors $E_1$ and $E_2$ are computed honestly from $T$, by using $\pi_T$ and the other components in the proof $\pi$;
    \item {\em Convergence verification}: $\mathcal{V}$ verifies if $E_1$ and $E_2$ satisfy the convergence condition (i.e., whether $\Delta W$ helps the model to reach convergence).  
\end{ditemize} 
Next, we discuss the {\bf Setup}, {\bf Certify} and {\bf Verify} protocols  respectively. Then we discuss how to deal with decimal and negative weights.

\subsection{Setup Protocol}
\label{sc:set}
Based on the public key, we define the following function for the data owner $\mathcal{O}$ to calculate a synopsis for each sample $(\vec{x}, y)$ in $T$. In particular, 
\begin{equation}
    d(\vec{x}, y)) = H(g^{x_1} || g^{x_2} || \dots || g^{x_m} || g^y),
    \label{eq:setup1}
\end{equation}
where $x_1$, $x_2$, $\dots$, $x_m$ are the features, $y$ is the label, and $g$ is the group generator. 

With the help the secret key $s$, $\mathcal{O}$ generates the signature $\gamma$ for $(\vec{x}, y)$ with 
    $\tau = d(\vec{x}, y))^s$.
Then instead of sharing the large amount of signatures with the verifier, $\mathcal{O}$  creates an aggregated signature 
    $\gamma = \pi_{i=1}^n \tau_i$,
where $\tau_i$ is the signature for the i-th sample in the training data $T$. 
Then $\gamma$ serves as a short signature of the whole dataset $T$.

\subsection{Certify Protocol}
\label{sc:certify}



To enable the verifier to verify $E_1$ and $E_2$ 
without access to the private samples $T=\{(\vec{x}, y)\}$, our Certify protocol construct a {\em proof} $\pi$ as following:
$\pi = \{\pi_E, \pi_W, \pi_T\}$,
where 
\begin{ditemize}
    \item $\pi_E=\{E_1, E_2\}$, i.e., $\pi_E$ stores the errors of the model. 
    \item $\pi_T= \{\{g^{x_i}\}, g^{y}|\forall (\vec{x}, y)\in T\}$, i.e., $\pi_T$ stores the digest of original data $\{\vec{x}\}$ and $\{y\}$. Storing the digest but not the original data is to due to the privacy concern in the outsourcing setting (Sec \ref{sc:framework}). 
    \item $\pi_W = \{\{\Delta w_{jk}^1\}, \{z_k^1\}, \{\hat{z}_k^1\}, g^{\delta^o}, \{\delta^L_k\}\}$, where $\Delta w_{jk}^1$ is the weight updated between the input and {\em first} hidden layer by one round of backpropagation after the model reaches convergence, $z_k^1$ and $\hat{z}_k^1$ are the weighted sum of the neuron $n_k^1$ (Eqn. \ref{eq:z}) at convergence and one round after convergence respectively, $\delta^o$ and $\{\delta^L_k\}$ are the error signals at output and the last hidden layer at convergence respectively. Intuitively, $\pi_W$ stores a subset of model outputs at the final two rounds (i.e., the round reaching convergence and one additional round afterwards). 
\end{ditemize}

\subsection{Verify Protocol}
\label{sc:verify}
The verification process consists of four steps: (1) authenticity verification of $\pi_T$; (2) one feedforward to verify the authenticity of $E_1$; (3) one backpropagation to update weights and  another feedforward to verify the authenticity of $E_2$; 
and (4) verification of convergence, i.e. if $|E_1-E_2|\leq \theta$, where $\theta$ is a pre-defined threshold for termination condition. 
Next, we discuss these steps in details. 

\noindent{\bf Step 1. Verification of $\pi_T$:} The verifier firstly verifies the authenticity of $\pi_T$, i.e., the digest of training samples. In particular, the verifier checks whether the following is true:
    $\Pi_{d(\vec{x}, y) \in \pi_T} e((\vec{x}, y), v) \stackrel{?}{=} e(\gamma, g)$,
where $d(\cdot)$ is the synopsis function (Eqn. (\ref{eq:setup1})), $v=g^s$ is a part of the public key, $\gamma$ is the aggregated signature provided by the data owner. If $\pi_T$ passes the verification, 
$\mathcal{V}$ is assured that the digests in $\pi_T$ are calculated from the intact dataset $T$.

\noindent{\bf Step 2. Verification of $E_1$:} 
First, the verifier $\mathcal{V}$ verifies if the weighted sum $\{z_k^1\}$ at the final round is correctly computed. Note that $\mathcal{V}$  is aware of $w_{ik}^1$. $\mathcal{V}$ also obtains $\{g^{x_i}\}$ and $\{z_k^1\}$ from $\pi_W$ in the proof. Then to verify the correctness of $\{z_k^1\}$, for each $z_k^1$, $\mathcal{V}$ checks if the following is true: 
\begin{equation}
    \Pi e(g^{x_i}, g^{w_{ik}^1}) \stackrel{?}{=} e(g, g)^{z_k^1}.
    \label{eq:verify1}
\end{equation}

Once $\mathcal{V}$ verifies the correctness of $\{z^1_k\}$, it calculates the activation of the hidden layers and thus the output $o$ (Eqns. (\ref{eq:a}) and (\ref{eq:forward_output})). 
Next, $\mathcal{V}$ checks if the following is true:  
\begin{equation}
    \Pi_{(\vec{x}, y)\in D} e(g^{y-o}, g^{y-o}) \stackrel{?}{=} e(g, g)^{2NE_1},
    \label{eq:verify2}
\end{equation}
where $g^{y - o}=g^{y}\ast g^{-o}$. Note that $g^{y}$ is included in the proof. $\mathcal{V}$  can compute $g^{-o}$ by using $o$  computed previously. 

\noindent{\bf Step 3. Verification of $E_2$:}
This step consists of five-substeps. 
The first four substeps verify the correctness of weight increment in the backpropagation process, including the verification of error signal at the output layer,  the verification of error signal at the last hidden layer, the verification of weight increments between all hidden layers, and verification of weight increments between the input and the first hidden layer. The last substep is to verify the authenticity of $E_2$ based on the updated weights. Next, we discuss  the details of these five substeps.

First, $\mathcal{V}$ verifies the correctness of $g^{\delta^o}$. 
Following Eqn. (\ref{eq:signal_1}), $\mathcal{V}$ can easily predict label $y$ with  $\delta^o$. Therefore, $\pi_W$ only includes $g^{\delta^o}$. $\mathcal{V}$ verifies the following: 
\begin{equation}
    \label{eq:verify4}
    e(g^{-o}g^{y}, g^{-\sigma'(z^o)}) \stackrel{?}{=} e(g, g^{\delta^o}),
\end{equation}
where $g^{-o}$ and $g^{-\sigma'(z^o)}$ are computed by $\mathcal{V}$, and $g^{y}$ and $g^{\delta^o}$ are from the proof.

Second, $\mathcal{V}$ verifies the correctness of $\delta_k^L$ (Eqn. (\ref{eq:signal_2})), i.e., the error signal on the $k$-th neuron on the last hidden layer, by checking if 
     $e(g^{w^o_{k}\sigma'(z_k^L)}, g^{\delta^o}) \stackrel{?}{=} e(g, g)^{\delta_k^L}$,
    \label{eq:verify5}
where $g^{w^o_{kj}\sigma'(z_k^L)}$ is computed by $\mathcal{V}$, and $\delta_k^L$ and $g^{\delta^o}$ are obtained from the proof. 

Third, $\mathcal{V}$ calculates the error signal of other hidden layers by following Eqn. (\ref{eq:signal_2}). 
Then with the knowledge of the activation on every hidden layer (by Step 2), $\mathcal{V}$ computes the derivatives of the weights (Eqn. \ref{eq:gradient}) on the hidden layers to update the weights between consecutive hidden layers (Equations \ref{eq:increment} - \ref{eq:update}).

Fourth, $\mathcal{V}$ verifies the weight increment between input and the first hidden layer. 
We must note that $\mathcal{V}$ cannot compute $\frac{\partial C}{\partial w_{jk}^{1}}$ (Eqn. (\ref{eq:gradient})) and $\Delta w_{jk}^1$ (Eqn. (\ref{eq:increment})) as it has no 
 access to the input feature $x_j$. Thus $\mathcal{V}$ obtains $\Delta w_{jk}^1$ from the proof $\pi$ and verifies its correctness by checking if the following is true: 
\begin{equation}
    \Pi_{(\vec{x}, y)\in D} e(g^{x_j}, g^{\eta \delta_k^1}) \stackrel{?}{=} e(g^{\Delta w_{jk}^1}, g^{-N}).
    \label{eq:verify3}
\end{equation}
Note that $g^{x_j}$ and $\Delta w_{jk}^1$ are included in the proof, and $g^{\eta \delta_k^1}$ and $g^{-N}$ are calculated by $\mathcal{V}$. After $\Delta w_{jk}^1$ is verified, 
$\mathcal{V}$ updates the weight by Eqn. (\ref{eq:update}).
Finally, $\mathcal{V}$ verifies $E_2$ by following the same procedure of Step 2 on the updated weights.

\noindent{\bf Step 4. Verification of convergence: } If $E_1$ and $E_2$ pass the authenticity verification, the verifier verifies the convergence of training by checking if $|E_1-E_2|\leq \theta$, i.e., it reaches the termination condition. 

We have the following theorem to show the security of \system. 
\begin{theorem}
\label{th:security}
The authentication protocols of \system is secure  (Definition \ref{def:security}). 
\end{theorem}
We omit the detailed proofs due to the limited space. Please refer to the extended version \cite{fullpaper} for it.
\subsection{Dealing with Decimal \&  Negative Values}
\label{sc:decimal}
One  weaknesses of bilinear pairing is that it cannot use decimal and negative values as the exponent in $g^e$. Therefore, the verification in Equations \ref{eq:verify1} - \ref{eq:verify3} cannot be performed easily. To address this problem, we extend the bilinear pairing protocol to handle decimal and negative values.

\noindent{\bf Decimal values.} 
We design a new method that conducts decimal arithmetic in an integer field without  accuracy loss. 
Consider the problem of checking if $b \ast c \stackrel{?}{=} e$, where $b$, $c$ and $e$ are three variables that may hold decimal values. Let $L_T$ be the maximum number of bits after the decimal point allowed for any value. 
We define a new operator $f(\cdot)$ where $f(x)=x \ast 2^{L_T}$. Obviously, $f(x)$ must be an integer. We pick two cyclic groups $G$ and $G_T$ of sufficiently large order $p$ such that $f(x)f(y) < Z_p$. Thus, we have $g^{f(x)}\in G$, and $e(g^{f(x)}, g^{f(y)})\in G_T$. 
To make the verification in Eqn. (\ref{eq:verify4}) applicable with decimal values, we check if $e(g^{f(b)}, g^{f(c)}) \stackrel{?}{=} e(g, g)^{f(e)}$. Obviously, if $e(g^{f(b)}, g^{f(c)}) = e(g, g)^{f(e)}$, it is natural that $b\ast c = e$. 
The verification in Eqn. (\ref{eq:verify1}), (\ref{eq:verify2}) and (\ref{eq:verify3}) is accomplished in the same way, except that the involved values should be raised by $2^{L_T}$ times. 

\noindent{\bf Negative values.} Equations  (\ref{eq:verify1} - \ref{eq:verify3}) check for a given pair of vectors $\vec{u}, \vec{v}$ of the same size, whether $\sum u_iv_i = z$. Note that the verification in Eqn. (\ref{eq:verify4}) can be viewed as a special form in which both $\vec{u}$ and $\vec{v}$ only include a single scalar value.
Also note that $u_i$, $v_i$ or $z$ may hold negative values.
Before we present our methods to deal with negative values, we first define an operator $[\cdot]$ such that $[x]=x\ mod\ p$.
\nop{
Since in the cyclic group $G$, we can only raise the generator $g$ to an exponent that is within the group $\mathbb{Z}_p$, where $p$ is a sufficiently large integer, for any value $x\in \mathbb{Z}$, we can only get $g^{[x]}$, and it is possible that $[x]\neq x$. 
}
Without loss of generality, we assume that for any $\sum u_iv_i=z$, $-p < u_i, v_i, z < p$. We have the following lemma.
\begin{lemma}
For any pair of vectors $\vec{u}, \vec{v}$ of the same size, and $z=\sum u_iv_i$, we have 
\[
\Big[\sum [u_i][v_i]\Big]=
\begin{cases}
z & \text{if } z\geq 0 \\
z+p & \text{otherwise.}
\end{cases}
\]
\label{lm:modular}
\end{lemma}


We omit the proof of Lemma \ref{lm:modular} due to limited space. Please refer to the extended version \cite{fullpaper} for it.
Following Lemma \ref{lm:modular}, we have Theorem \ref{thm:negative} to verify vector dot product operation in case of negative values based on bilinear pairing.
\begin{theorem}
\label{thm:negative}
To verify $\sum u_iv_i \stackrel{?}{=} z$, it is equivalent to checking if 
\begin{equation}
\Pi e(g^{[u_i]}, g^{[v_i]}) \stackrel{?}{=} 
\begin{cases}
e(g, g)^z & \text{if } z\geq 0 \\
e(g, g)^{(z+p)} & \text{otherwise.}
\end{cases}
\label{eq:negative}
\end{equation}
\end{theorem}
We omit the proof due to the simplicity and include it in the extended version \cite{fullpaper}.
\nop{
 The error $E_1$ cannot be negative in Eqn. (\ref{eq:verify2}). Therefore, we can perform the verification of weighted sum by following Eqn. (\ref{eq:negative}), even though $y_j-o_j$ might be negative values.

 Eqn. (\ref{eq:verify1}) is different from Eqn. (\ref{eq:verify2}) in that the weighted sum result $z_k^1$ can be either positive or negative. This difference also holds in Eqn. (\ref{eq:verify4} - \ref{eq:verify3}).
 }
 Next, we focus on Eqn. (\ref{eq:verify1}) and discuss our method to handle negative values. First, based on Lemma \ref{lm:modular}, we can see that for any $x_i$ and $w_{ik}^1$, if $x_iw_{ik}^1\geq 0$, then $[x_i][w_{ik}^1]=x_iw_{ik}^1$; otherwise, $[x_i][w_{ik}^1]=x_iw_{ik}^1+p$. Therefore, to prove $z_k^1=\sum x_i w_{ik}^1$, the server includes a flag $sign_i$ for each $x_i$ in the proof, where 
\[
sign_i = 
\begin{cases}
+ & \text{if } x_i\geq 0 \\
- & \text{otherwise}.
\end{cases}
\]
Meanwhile, for each $z_k^1$, the server prepares two values $p_k^1=\sum_{i: x_iw_{ik}^1\geq 0} x_iw_{ik}^1$ and $n_k^1=\sum_{i: x_iw_{ik}^1< 0} x_iw_{ik}^1$, and includes them in the proof.

In the verification phase, since the client is aware of $w_{ik}^1$, with the knowledge of $sign_i$ in the proof, it can tell if $x_iw_{ik}^1\geq 0$ or not. So the client first verifies if 
\[
\Pi_{i: x_iw_{ik}^1\geq 0} e(g^{[x_i]}, g^{[w_{ik}^1]}) \stackrel{?}{=} e(g,g)^{p_k^1}, \Pi_{i: x_iw_{ik}^1< 0} e(g^{[x_i]}, g^{[w_{ik}^1]}) \stackrel{?}{=} e(g,g)^{n_k^1+p},
\]
where $g^{[x_i]}$ is included in the proof, and $g^{[w_{ik}^1]}$ is computed by the client.
Next, the client checks if $p_k^1+n_k^1\stackrel{?}{=}z_k^1$.

\section{Experiments}
\label{sc:exp}
\subsection{Setup}
\label{sc:setup}
{\bf Hardware \& Platform.}
We implement \system in C++. We use the implementation of bilinear mapping from PBC library\footnote{https://crypto.stanford.edu/pbc/.}. The DNN model is implemented in Python on TensorFlow.
We simulate the server on a computer of 2.10GHz CPU, 48 cores and 128GB RAM, and the data owner and the verifier on 2 computers of 2.7GHz Intel CPU and 8GB RAM respectively. 

{\bf Datasets.}  We use the following four datasets that are of different data types: (1) {\bf MNIST} dataset that contains 60,000 image  samples and 784 features; (2)  {\bf TIMIT} dataset that contains 4,620 samples of broadband recordings and 100 features; 
(3) {\bf ADULT} dataset that includes 45,222 records and 14 features; and 
(4) {\bf HOSPITAL} dataset that contains 230,000 records and 33 features. 


\nop{
\begin{table}[ht!]
\centering
\begin{tabular}{|c|c|c|c|c|}
\hline
Dataset        & \# of training & \# of testing & \# of  & \# of  \\
& samples & samples & features & classes \\ \hline
MNIST          & 60,000  & 6,000 & 784    & 10          \\ \hline
TIMIT & 4,620 & 462 &  100      & 61          \\ \hline
ADULT & 30,000 & 3,000 &  14      & 2          \\ \hline
HOSPITAL & 230,000 & 10,000 &  33      & 20          \\ \hline
\end{tabular}
\caption{\label{tb:dataset}Description of datasets}
\end{table}
}

\nop{
\begin{table}[!htbp]
    \centering
    \begin{tabular}{|c|c|c|c|c|}
\hline
Minibatch Size         & 200    & 300    & 400    & 500    \\ \hline
B-VERIFL & 233.28 & 348.15 & 462.36 & 574.28 \\ \hline
O-VERIFL & 2.24 & 3.18 & 4.12 & 5.03 \\ \hline
\end{tabular}
    \caption{\small{Proof construction time (seconds)}}
\label{tb:proofctime}
\end{table}
}


{\bf Neural network architecture.} We train a DNN with four fully connected hidden layers for the MNIST, ADULT and HOSPITAL datasets. We vary the number of neurons on each hidden layer from 10 to 50, and the number of parameters from 20,000 to 100,000. We apply sigmoid function on each layer, except for the output layer, where we apply softmax function instead. We optimize the network by using gradient descent with the learning rate $\eta=0.1$. By default, the minibatch size is $100$. We use the same DNN structure for the TIMIT dataset with ReLU as the activation function.  

\nop{
{\bf Simulation of Server's Misbehavior.} 
We experimentally evaluate the efficacy of \system~ by simulating three types of misbehavior of the server, and checking if \system~ can detect them.
We simulated the Byzantine failures of participants by randomly choosing 1\% neurons and replacing the output of these neurons with random values. 
We generated three types of wrong local model updates: (1) the participant sends the wrong error $E_1$ with the proof constructed from correct $E_1$; (2) the participant sends wrong $E_1$ with the proof constructed from wrong $E_1$; (3) the participant sends correct $E_1$ and wrong $E_2$. 
}

{\bf Basic and optimized versions of \system.} We implement two versions of \system: (1) Basic approach ({\bf B-VERIDL}): the proof of model updates is generated for every single input example $(\vec{x}, y)$; and (2) Optimized approach ({\bf O-VERIDL}): the proof is generated for every unique value in the input $\{(\vec{x}, y)\}$. 



{\bf Existing verification approaches for comparison.} We compare the performance of \system with two alternative approaches:
(1) {\bf $C_1$. Homomorphic encryption (LHE) vs. bilinear mapping:}  When generating the proof, we use LHE to encrypt the plaintext values in the proof instead of bilinear mapping; (2)  {\bf $C_2$. Result verification vs. re-computation of model updates by privacy-preserving DL:}  The server encrypts the private input samples with homomorphic encryption. The verifeir executes the learning process on the encrypted training data, and compares the computed results with the server's returned updates. 
For both comparisons, we use three different implementations of HE. The first implementation is the Brakerski-Gentry-Vaikuntanathan (BGV) scheme provided by HElib library\footnote{https://github.com/shaih/HElib.}. The second implementation is built upon the PALISADE library\footnote{https://git.njit.edu/palisade/PALISADE/wikis/home} that uses primitives of lattice-based cryptography for implementation of HE. 
The last one is built upon the Microsoft SEAL project \cite{sealcrypto}, which provides a programming interface to lightweight homomorphic encryption.

\subsection{Efficiency of \system}

\begin{figure}[!htbp]
\centering
\begin{tabular}{cc}
    \includegraphics[width=0.4\textwidth]{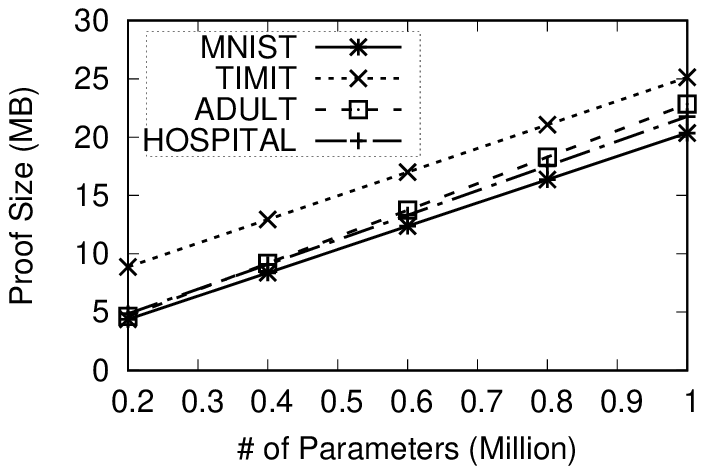}
    &
    \includegraphics[width=0.4\textwidth]{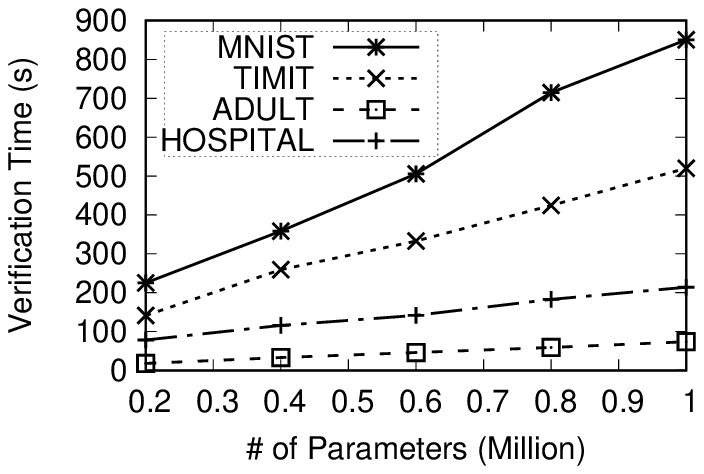}
    \\
    {\small (a) Proof size}
     &
     {\small (b) Verification time}
	\end{tabular}
  \caption{\label{fig:Dperformance5} Performance of \system (minibatch size 100)}
\end{figure}

\begin{figure*}[!htbp]
\centering
\begin{tabular}{cc}
    \includegraphics[width=0.4\textwidth]{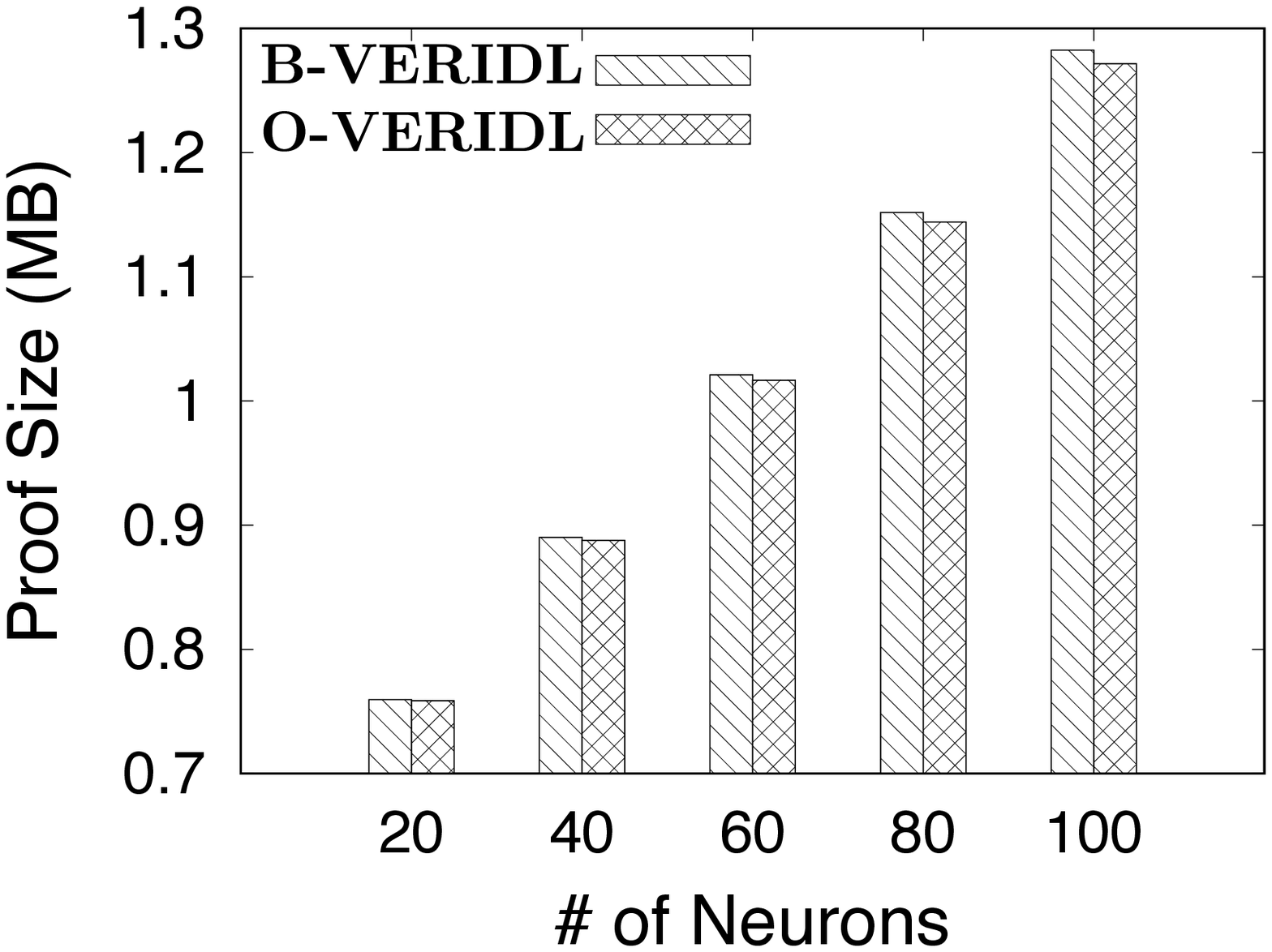}
    &
    \includegraphics[width=0.4\textwidth]{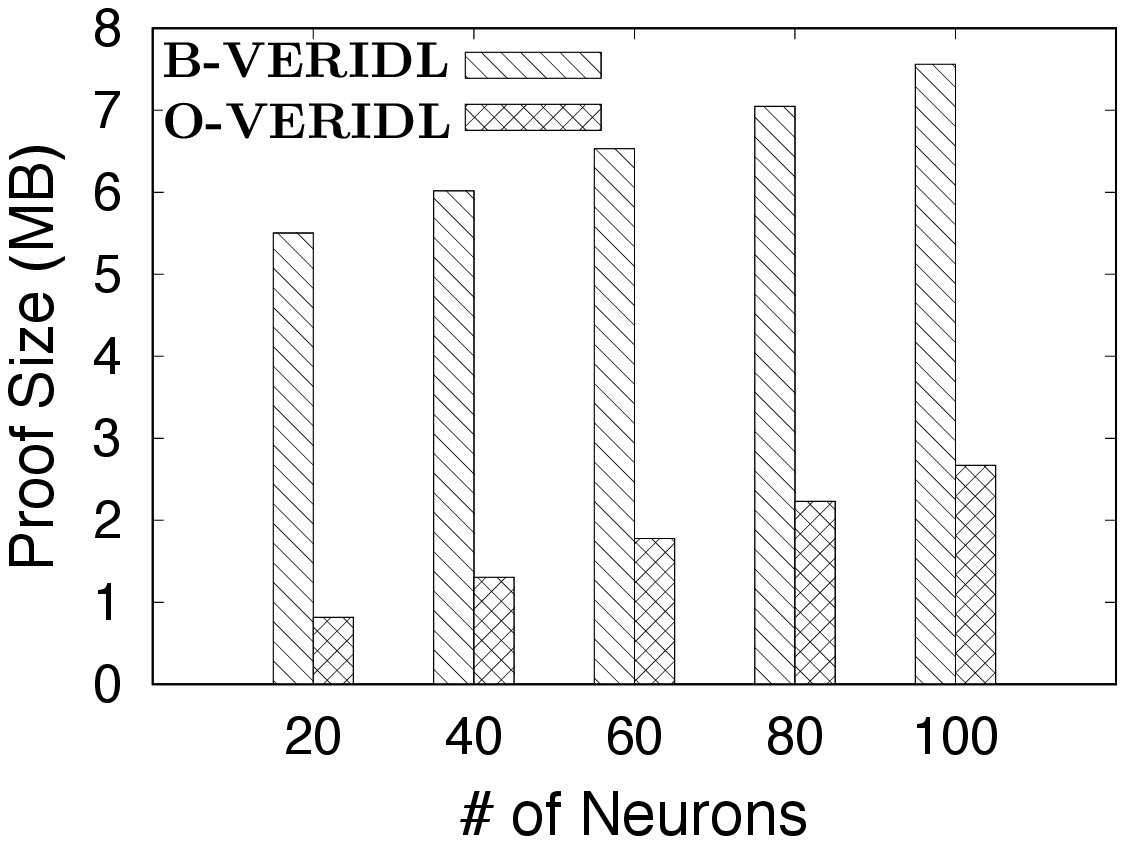}
    \\
     {\small (a)  MNIST dataset}
      &
      {\small (b) TIMIT dataset}
	\end{tabular}
  \caption{\label{fig:Dperformance2} Proof size}
\end{figure*}

{\bf Proof size.} The results of proof size of \system on four datasets, with various number of neurons at each hidden layer, are shown in Figure \ref{fig:Dperformance5} (a). 
In all settings, the proof size is small (never exceeds 25MB even with one million parameters). This demonstrates the benefit of using bilinear pairing for proof construction.   
Second, we observe a linear increase in the proof size with the growth of the number of parameters. This is because the dominant components of proof size is the size of $\{\Delta w_{jk}^1\}$, $\{z_k^1\}$ and $\{\hat{z}_k^1\}$, which grows with the number of parameters. 

{\bf Verification time.} The results of verification time on all four datasets are shown in Figure \ref{fig:Dperformance5} (b). 
First, the verification time is affordable even on the datasets of large sizes. Second, the verification time grows linearly with the number of hyperparameters. The reason is that the number of neurons on the first hidden layer increases linearly with the growth of parameters in the neuron network,  while the verification time linearly depends on the input dimension and the number of neurons in the first hidden layer. 

\noindent{\bf B-\system VS. O-\system.} We compare the performance of the basic and optimized versions of \system. Figure \ref{fig:Dperformance2} demonstrates the proof size of B-\system and O-\system with various number of neurons at each hidden layer in the DNN model. 
In general, the proof size is small (less than 1.3MB and 8MB for MNIST and TIMIT datasets respectively). 
Furthermore, the proof size of O-VERIDL can be  smaller than B-VERIDL; it is 20\% - 26\% of the size by B-VERIDL on TIMIT dataset. This demonstrates the advantage of O-VERIDL. 
The results also show that the proof size of both O-VERIDL and B-VERIDL gradually rises when the number of neurons increases. 
However, the growth is moderate. This shows that \system can be scaled to large DNNs.

\begin{figure*}[t!]
\centering
\begin{tabular}{@{}c@{}c@{}c@{}c@{}c@{}}
    \includegraphics[width=0.4\textwidth]{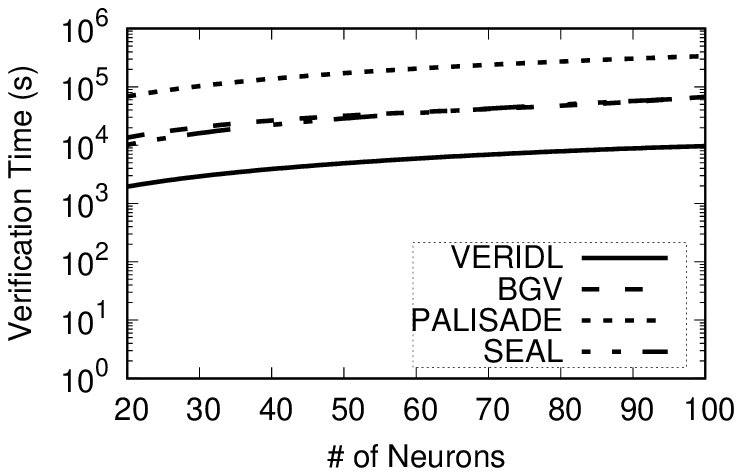}
    &
    \includegraphics[width=0.4\textwidth]{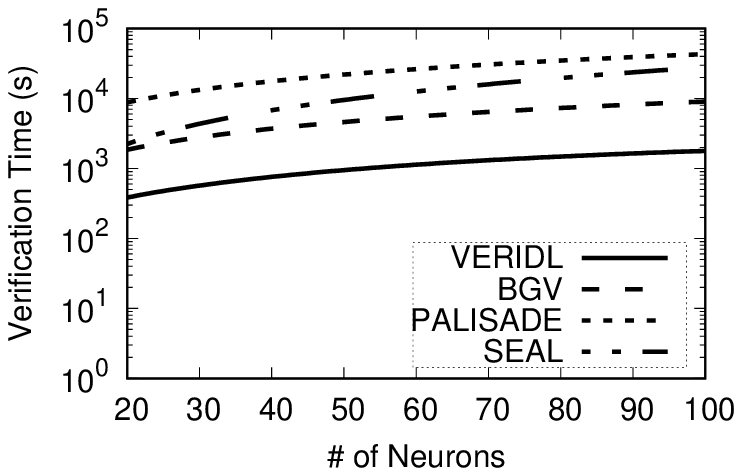}
    \\
    {\small (a) MNIST dataset}
     &
     {\small (b) TIMIT dataset}
	\end{tabular}
  \caption{\label{fig:Dperformance3} Verification time (minibatch size 100)} 
\end{figure*}

{\bf Comparison with existing approaches.}
We evaluate the verification time of different proof generation methods (defined by the comparison $C_1$ in Section \ref{sc:setup}) for various numbers of neurons on all four datasets, and report the results of MNIST and TIMIT datasets in Figures \ref{fig:Dperformance3}. The results on ADULT and HOSPITAL datasets are similar; we omit them due to the limited space. 
we observe that for all four datasets, \system (using bilinear mapping) is more efficient than using HE (i.e., BGV, PALISADE and SEAL) in the proof. Thus bilinear mapping is a good choice as it enables the same function over ciphertext with cheaper cost. 
Besides, the time performance of both \system and HE increases when the number of neurons in the network grows. This is expected as it takes more time to verify a more complex neural network. 
We also notice that all approaches take longer time on the MNIST dataset than the other datasets. This is because the MNIST dataset includes more features than the other datasets; it takes more time to complete the verification in Equations \ref{eq:verify1} - \ref{eq:verify3}.  


\subsection{Verification vs. Re-computation of Model Updates}
\begin{figure}[t!]
\begin{center}
\begin{tabular}{cc}
    \includegraphics[width=0.4\textwidth]{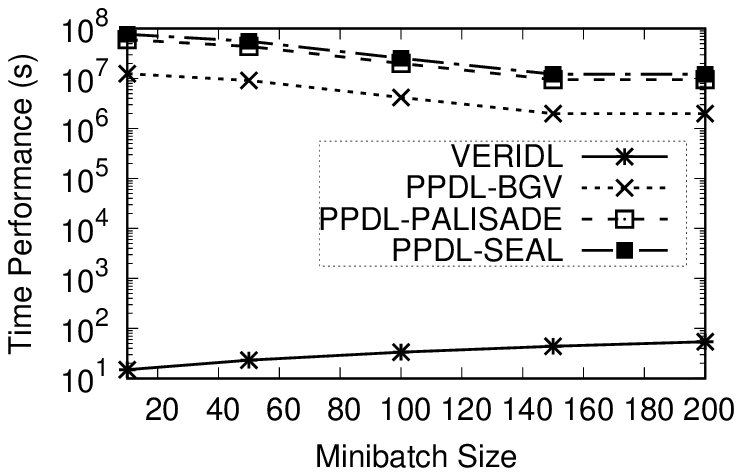}
    &
    \includegraphics[width=0.4\textwidth]{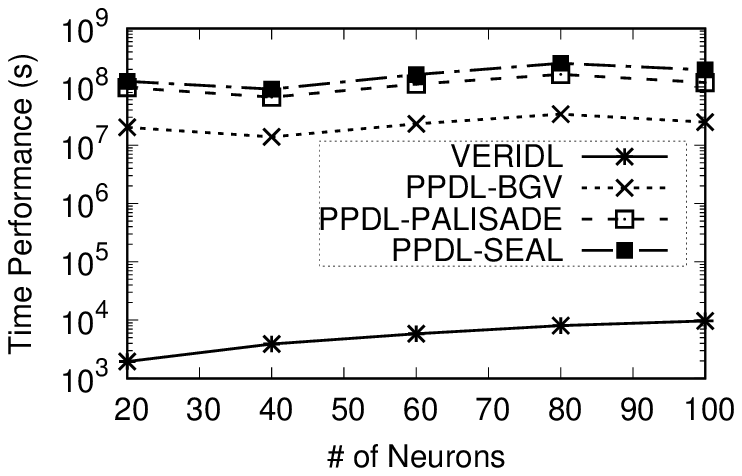}
     \\
     (a) Minibatch size
     &
     (b) \# of neurons
	\end{tabular}
  \caption{\label{fig:Dperformance4}  Verification vs. re-computation of model updates}
 \end{center} 
\end{figure}

We perform the comparison $C_2$ (defined in Sec. \ref{sc:setup}) by implementing the three HE-based privacy-preserving deep learning (PPDL) approaches  
 \cite{gilad2016cryptonets,hesamifard2017cryptodl,sealcrypto} and comparing the performance of \system with them. To be consistent with \cite{gilad2016cryptonets,hesamifard2017cryptodl}, we use the approximated ReLU as the activation function due to the fact that HE only supports low degree polynomials. 
Figure \ref{fig:Dperformance4} shows the comparison results. 
In Figure \ref{fig:Dperformance4} (a), we observe that \system is faster than the three PPDL methods by more than three orders of magnitude. 
An interesting observation is that \system and PPDL take opposite pattern of time performance when the minibatch size grows. The main reason behind the opposite pattern is that when the minibatch size grows, \system has to verify $E_1$ and $E_2$ from more input samples (thus takes longer time), while PPDL needs fewer epochs to reach convergence (thus takes less time).
Figure \ref{fig:Dperformance4} (b) shows the impact of the number of neurons on  the time performance of both \system and PPDL. Again, \system wins the three PPDL methods by at least three orders of magnitude. This demonstrates that \system is more efficient than verification by PPDL. 

\subsection{Robustness of Verification}
\label{sc:defense}
\nop{
\begin{figure}[t!]
\begin{tabular}{cc}
    \includegraphics[width=0.4\textwidth]{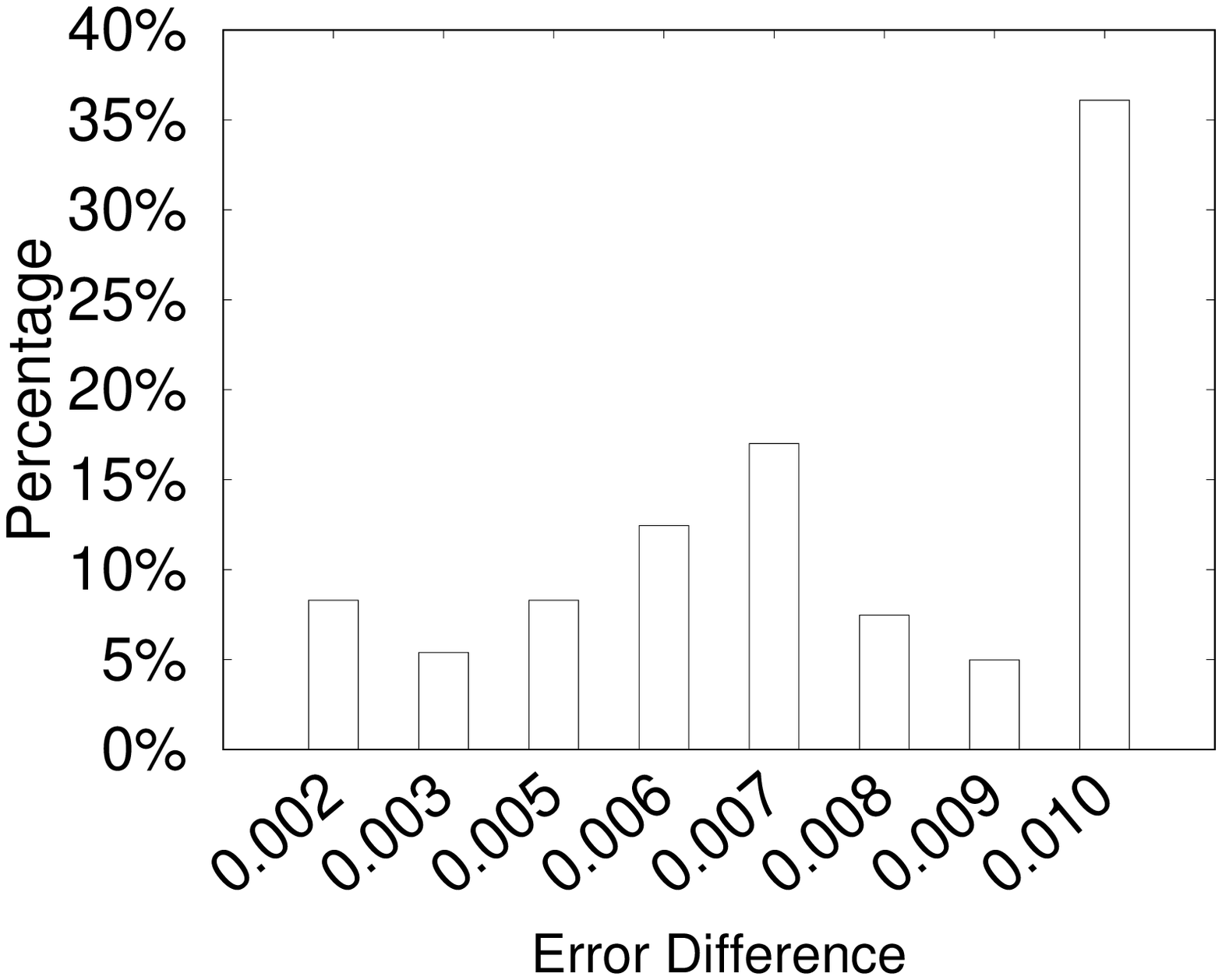}
    &
    \includegraphics[width=0.4\textwidth]{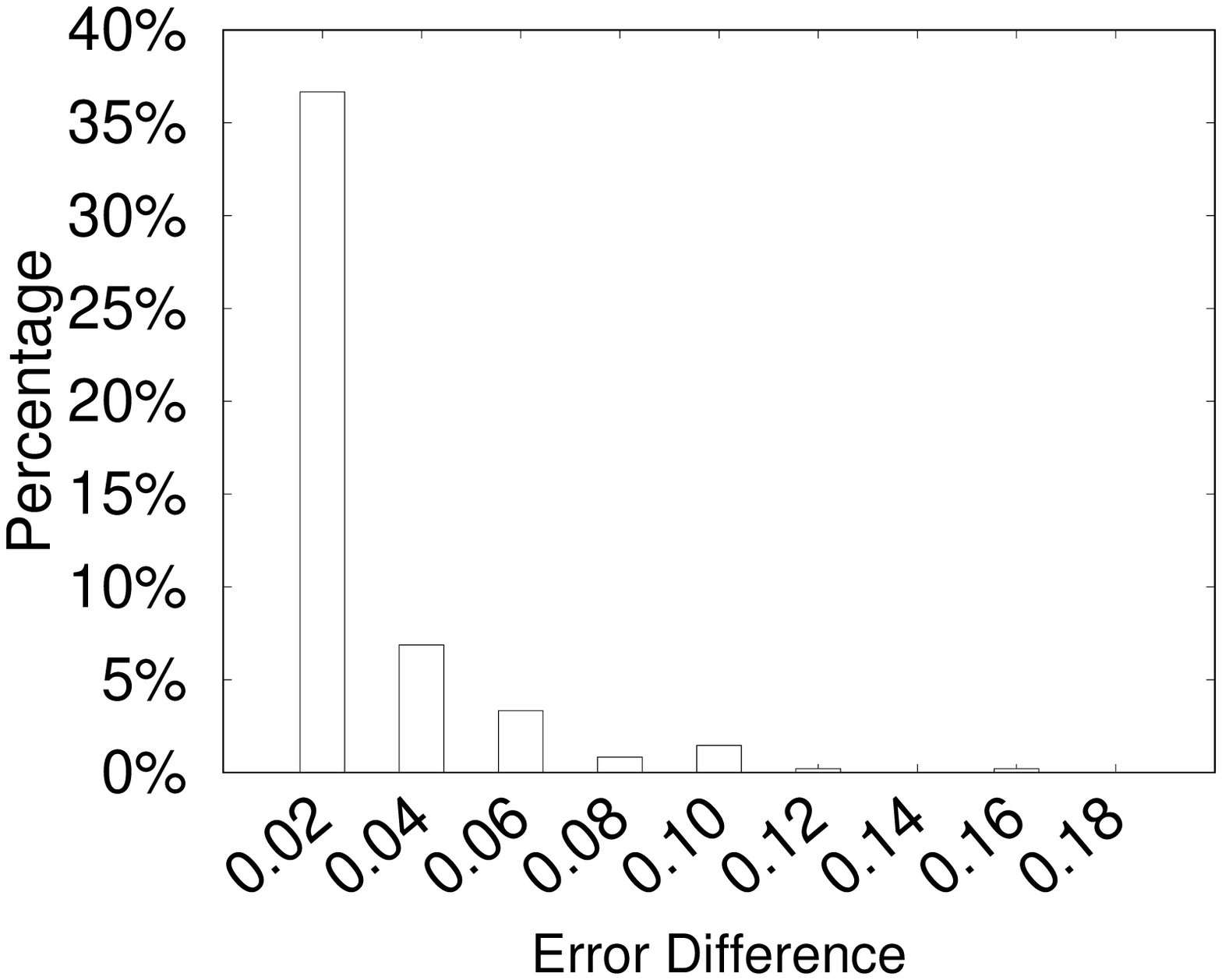}
     \\
     (a) Low-precision floating points attack
     &
     (b) Network pruning attack
	\end{tabular}
  \caption{\label{fig:Dperformance10} Distribution of difference in the errors of compressed and original models by the compression attack}
\end{figure}
}

To measure the robustness of \system, we implement two types of server's misbehavior, namely {\em Byzantine failures} and  {\em model compression attack}, 
and evaluate if \system\ can catch the incorrect model updates by these misbehavior. 
\nop{
\begin{ditemize}
    \item We implement two types of model compression attacks, namely the {\em network pruning attack} and the {\em low-precision floating points attack}. For the network pruning attack, we prune 10\% - 25\%  edges from the neural network. For the low-precision floating points attack, we use the weights of 8-bit and 16-bit floating points instead of the  32-bit floating points. 
    \item For local model poisoning attack, we use the implementation package\footnote{\url{https://github.com/inspire-group/ModelPoisoning}} which accompanies the model poisoning attack in  \cite{bhagoji2019analyzing}. We run the code on a NVIDIA RTX 2080 Ti GPU. 
\end{ditemize} 
}

\noindent{\bf Byzantine failure.}
We simulate the Byzantine failure by randomly choosing 1\% neurons and replacing the output of these neurons with random values. We generate three types of wrong model updates: (1) the server sends the wrong error $E_1$ with the proof constructed from correct $E_1$; (2) the server sends wrong $E_1$ with the proof constructed from wrong $E_1$; (3) the server sends correct $E_1$ and wrong $E_2$. 
Our empirical results demonstrate that \system~ caught all wrong model updates by these Byzantine failures with 100\% guarantee. 

\noindent{\bf Model compression attack.} 
The attack compresses a trained DNN network with small accuracy degradation \cite{courbariaux2014training,gong2014compressing}. 
To simulate the attack, we setup a fully-connected network with two  hidden layers and sigmoid activation function. The model parameters are set by randomly generating 32-bits weights. We use ADULT dataset as the input.
We simulate two types of model compression attacks: (1) the {\em low-precision floating points attack} that truncates the initial weights to 8-bits and 16-bits respectively and train the truncated weights; and (2) the {\em network pruning attack} that randomly selects 10\% - 25\% weights to drop out during training. For both attacks, we run 50 times and calculate the absolute difference between the error $E_1'$ computed from the compressed model and the error $E_1$ of the correct model.
From the results, we observe that the error difference produced by the low-precision attack is relatively high (with a 35\% chance of less than or equal to 0.02), and can be as large as 0.2. While the error differences of the network pruning attack are all between 0.002 and 0.01.
In all cases, we have $|E_1' - E_1| \geq 10^{-9}$. We omit the results due to the limited space. 
We must note that given the DNN model is a 32-bit system, \system can determine that $E_1'\neq E_1$  as long as $|E_1' - E_1|\geq 10^{-9}$. 
Therefore, \system can detect the incorrect model updates by both network compression attacks, even though the attacker may forge the proof of $E_1$ to make $E_1'$ pass the verification.



\nop{
\noindent{\bf Model poisoning attack.} 
This attack aims at compromising the returned model in certain direction\cite{fang2019local,bhagoji2018model}. 
\cite{bhagoji2019analyzing} designs an attack in which the malicious server manipulates model updates by using explicit boosting to achieve targeted mis-classification.
In our experiments, we firstly calculate the correct model updates. It turned out that for all the settings,  the correct error $E_1$ of the model is in the range of $[0.19, 0.46]$. Then we craft the model updates by using the code\footnote{\url{https://github.com/inspire-group/ModelPoisoning}} of the the basic targeted model poisoning attack in  \cite{bhagoji2019analyzing}.  We measure the error $E_1'$ of the crafted model updates. Our results show that $E_1'$ is significantly different from $E_1$; it can be as large as $860937.75$. 
Furthermore, the incorrect model updates fail to make the model converge on the original training data $D$.
 We simulate two possible scenarios that the adversary may try to escape the verification. In the first scenario, the adversary forges the proof by setting $\pi_E=\{E_1, E_2\}$ (i.e., the correct error).  \system is able to detect that the proof is inconsistent with the model.
In the second scenario, the adversary constructs the proof of $\{E_1' ,E_2'\}$, and sends the proof together with the wrong model updates. Again, \system detects the wrong updates as the incorrect model updates cannot make the model converge. We omit the details of these experiments due to limited space. 
}

\nop{
\section{Discussion}
\label{sc:disc}

\nop{
\noindent{\bf Data integrity.} 
\system assumes that the integrity of the local data from participants has been verified. 
Existing researches provide many data integrity verification approaches. For example, prior to the start of FL, the participant can use one-way hash function (e.g., SHA-3) to generate a constant-size digest of the local data, so that the server can check data integrity by applying the same hash function over the data sent by the participant and comparing the result with it. 
}

{\bf Verification of model updates over polluted data.} \system assumes that the data used for training of local models is genuine and unpolluted. However, the malicious participant may pollute  the data to produce incorrect local model updates. Some examples of such pollution attacks include: (1) DNN backdoor attacks 
\cite{bagdasaryan2018backdoor,liu2017trojaning} that train the local DNN models with a random subset of training data points are stamped with the trigger pattern and their labels are modified into the target label;  (2) the data poisoning attack that modifies the inputs (e.g., by inserting adversarial examples) 
\cite{biggio2012poisoning,jagielski2018manipulating}; and 
(3) the targeted model poisoning attack \cite{bhagoji2019analyzing} that injects the auxiliary targets and their labels into training data. 
Formally, let $\pi'$ and $\Delta L'$ be the proof and the (incorrect) model updates computed from the polluted data $D'$. 
If the adversary follows the {\bf Certify} protocol of \system honestly to  construct $\pi'$ from $\Delta L'$, and uses $\pi'$ as the proof to accompany $\Delta L'$, \system cannot catch $\Delta L'$ as  wrong. This is one of the biggest limitations of \system. 
One possible solution is to utilize the existing defense mechanisms against adversarial examples (e.g.,  \cite{meng2017magnet,pang2018towards}) for verification of data integrity (i.e., if data has been tampered), and our \system for verification of model integrity. 

\noindent{\bf Inference-based privacy attack against \system.} In the proof that is generated by \system, the same plaintext values will have the same signatures. Due to this fact, the signatures in the proof can disclose the frequency of the data. If the server is aware of the frequency distribution in the participant's data, he/she can recover a part of the private instances. 
To block the inference channel, one solution is to allow the participants to add small noise to the local training data, and execute optimization and generate signatures from the noisy data. In deep learning, augmenting training data with small noise can dramatically reduce the generalization error \cite{poole2014analyzing}. Therefore, the accuracy of the aggregated global model is preserved. 

}
\section{Related Work}
\label{sc:related}
{\em Verified artificial intelligence (AI)} \cite{seshia2016towards} aims to design  AI-based systems that are provably correct with respect to mathematically-specified requirements. 
{\em DeepXplore} \cite{pei2017deepxplore} provides an automated white-box testing system for DL systems. It generates the corner cases where DL systems may generate unexpected or incorrect behaviors. 
SafetyNet \cite{ghodsi2017safetynets} provides a protocol that verifies the execution of DL on an untrusted cloud. The verification protocol is built on top of the {\em interactive proof} (IP) protocol and arithmetic circuits. The protocol only can be adapted to the DNNs expressed as arithmetic circuits. This places a few restrictions on DNNs, e.g., the activation functions must be polynomials with integer coefficients, which disables the activation functions that are commonly used in DNNs such as ReLU, sigmoid and softmax.
Recent advances in zero-knowledge (ZK) proofs significantly reduce the verification and communication costs, and make the approach more practical to verify delegated computations in public \cite{yang2021quicksilver}. ZEN \cite{feng2021zen} is the first ZK-based protocol that enables privacy-preserving and verifiable inferences for DNNs. However, ZEN only allows ReLU activation functions.
We remove such strict assumption. 
 {\em VeriDeep} \cite{he2018verideep} generates a few minimally transformed inputs named {\em sensitive samples} as fingerprints of DNN models. If the adversary makes changes to a small portion of the model parameters, the outputs of the sensitive samples from the model also change. However, VeriDeep only can provide a probabilistic correctness guarantee. 
\section{Conclusion and Future Work}
\label{sc:conclusion}
In this paper, we design \system, an authentication framework that supports efficient integrity verification of DNN models in the DLaaS paradigm. \system extends the existing bilinear grouping technique significantly to handle the verification over DNN models. The experiments demonstrate that \system can verify the correctness of the model updates with cheap overhead.

While \system provides a deterministic guarantee by verifying the output of {\em all} neurons in DNN, generating the proof for such verification is time costly. Thus an interesting direction to explore in the future is to design an alternative {\em probabilistic} verification method that provides high guarantee (e.g., with 95\% certainty) but with much cheaper verification overhead. 


\bibliographystyle{plain}
{
\bibliography{bib}

\begin{thebibliography}{10}

\bibitem{bengio2009learning}
Yoshua Bengio.
\newblock Learning deep architectures for ai.
\newblock {\em Foundations and trends in Machine Learning}, 2(1):1--127, 2009.

\bibitem{courbariaux2014training}
Matthieu Courbariaux, Yoshua Bengio, and Jean-Pierre David.
\newblock Training deep neural networks with low precision multiplications.
\newblock {\em arXiv preprint arXiv:1412.7024}, 2014.

\bibitem{fullpaper}
Boxiang Dong, Bo~Zhang, and Hui~(Wendy) Wang.
\newblock Veridl: Integrity verification of outsourced deep learning services
  (full version).
\newblock {\em arXiv preprint arXiv:2007.11115}, 2021.

\bibitem{feng2021zen}
Boyuan Feng, Lianke Qin, Zhenfei Zhang, Yufei Ding, and Shumo Chu.
\newblock Zen: Efficient zero-knowledge proofs for neural networks.
\newblock {\em IACR Cryptol. ePrint Arch.}, 2021:87, 2021.

\bibitem{ghodsi2017safetynets}
Zahra Ghodsi, Tianyu Gu, and Siddharth Garg.
\newblock Safetynets: Verifiable execution of deep neural networks on an
  untrusted cloud.
\newblock In {\em Advances in Neural Information Processing Systems}, pages
  4675--4684, 2017.

\bibitem{gilad2016cryptonets}
Ran Gilad-Bachrach, Nathan Dowlin, Kim Laine, Kristin Lauter, Michael Naehrig,
  and John Wernsing.
\newblock Cryptonets: Applying neural networks to encrypted data with high
  throughput and accuracy.
\newblock In {\em International Conference on Machine Learning}, pages
  201--210, 2016.

\bibitem{gong2014compressing}
Yunchao Gong, Liu Liu, Ming Yang, and Lubomir Bourdev.
\newblock Compressing deep convolutional networks using vector quantization.
\newblock {\em arXiv preprint arXiv:1412.6115}, 2014.

\bibitem{he2018verideep}
Zecheng He, Tianwei Zhang, and Ruby~B Lee.
\newblock Verideep: Verifying integrity of deep neural networks through
  sensitive-sample fingerprinting.
\newblock {\em arXiv preprint arXiv:1808.03277}, 2018.

\bibitem{hesamifard2017cryptodl}
Ehsan Hesamifard, Hassan Takabi, and Mehdi Ghasemi.
\newblock Cryptodl: Deep neural networks over encrypted data.
\newblock {\em arXiv preprint arXiv:1711.05189}, 2017.

\bibitem{lecun2015deep}
Yann LeCun, Yoshua Bengio, and Geoffrey Hinton.
\newblock Deep learning.
\newblock {\em nature}, 521(7553):436, 2015.

\bibitem{papamanthou2011optimal}
Charalampos Papamanthou, Roberto Tamassia, and Nikos Triandopoulos.
\newblock Optimal verification of operations on dynamic sets.
\newblock In {\em Annual Cryptology Conference}, pages 91--110, 2011.

\bibitem{pei2017deepxplore}
Kexin Pei, Yinzhi Cao, Junfeng Yang, and Suman Jana.
\newblock Deepxplore: Automated whitebox testing of deep learning systems.
\newblock In {\em Proceedings of the 26th Symposium on Operating Systems
  Principles}, pages 1--18. ACM, 2017.

\bibitem{sealcrypto}
{M}icrosoft {SEAL} (release 3.5).
\newblock \url{https://github.com/Microsoft/SEAL}, April 2020.
\newblock Microsoft Research, Redmond, WA.

\bibitem{seshia2016towards}
Sanjit~A Seshia, Dorsa Sadigh, and S~Shankar Sastry.
\newblock Towards verified artificial intelligence.
\newblock {\em arXiv preprint arXiv:1606.08514}, 2016.

\bibitem{yang2021quicksilver}
Kang Yang, Pratik Sarkar, Chenkai Weng, and Xiao Wang.
\newblock Quicksilver: Efficient and affordable zero-knowledge proofs for
  circuits and polynomials over any field.
\newblock {\em IACR Cryptol. ePrint Arch.}, 2021:76, 2021.

\end{thebibliography}
}

\nop{
\section{Appendix}
\subsection{Proof of Theorem 4.1 (Security Analysis)}


\nop{
\begin{theorem}
\label{th:security}
The authentication protocols of \system is secure  (Definition \ref{def:security}). 
\end{theorem}
}
\noindent{\em Proof.} 
Let $\Delta L$ be the correct local model update, and $\pi=\{\pi_E, \pi_D, \pi_W\}$ the correct proof of $\Delta L$.
Suppose an adversary $\mathbf{Adv}$ returns an incorrect local update $\Delta L'\neq \Delta L$ and the proof $\pi'=\{\pi_E', \pi_D', \pi_W'\}$. We assume that $\mathbf{Adv}$ is fully aware of the details of our authentication method. Thus it will try to generate $\pi'$ intelligently, trying to escape from the authentication. 

According to Formula (\ref{eq:verify3}), $\pi_W$ and $\pi_D$ must be consistent (i.e., either $\pi_D'=\pi_D$  and $\pi_W'=\pi_W$  or $\pi_D'\neq\pi_D$ and $\pi_W'\neq\pi_W$) in order to pass the verification. Therefore, there are four types of possible ways to generate the proof $\pi'$ by $\mathbf{Adv}$: 
\vspace{-0.05in}
\begin{itemize}
    \item $\pi_E'= \pi_E$, $\pi_D'=\pi_D$, and $\pi_W'=\pi_W$ (i.e., correct errors and correct proof); 
    \item $\pi_E'= \pi_E$, $\pi_D'\neq \pi_D$, and $\pi_W'\neq \pi_W$ (i.e., correct errors and incorrect proof); 
    \item $\pi_E'\neq \pi_E, \pi_D'=\pi_D, \pi_W'=\pi_W$ (i.e., incorrect errors and correct proof); 
    \item $\pi_E'\neq \pi_E, \pi_D'\neq \pi_D, \pi_W'\neq \pi_W$ (i.e., incorrect errors and incorrect proofs).
\end{itemize} 
\vspace{-0.05in}
Next, we discuss these behaviors case by case, and show that for each case, the incorrect local update cannot escape from verification (Def. \ref{def:security}).

\noindent{\em Case 1: $\pi_E'= \pi_E$, $\pi_D'=\pi_D$, and $\pi_W'=\pi_W$}. For this case, the verification by Formula (\ref{eq:verify1}) can only pass with negligible probability, i.e., $\epsilon(\ell)$, since $z_k^1$ is generated by using the weights with correct update. It cannot match with $z_k^1$ by using incorrect weights. 

\noindent{\em Case 2: $\pi_E'= \pi_E$, $\pi_D'\neq \pi_D$, and $\pi_W'\neq \pi_W$}. This case can be detected due to the inconsistency between the correct error $\pi_E=\{E_1, E_2\}$ and the incorrect model. In particular, it can be detected by Formula (\ref{eq:verify2}). This is because $E_1 \neq \frac{1}{N}\sum_{(\vec{x}, y) \in D} C(\vec{x}, y; W+\Delta L')$. Hence we can verify the correctness of $E_1$ for any crafted model updates. 

\noindent{\em Case 3: $\pi_E'\neq \pi_E, \pi_D'=\pi_D$, and $\pi_W'=\pi_W$}. The adversary $\mathbf{Adv}$ returns incorrect $\pi_E'=\{E_1', E_2'\}$, where either $E_1'\neq E_1$ or $E_2'\neq E_2$, and the correct proof $\pi_D$ constructed from the right $E_1$ and $E_2$. Then $\mathbf{Adv}$ can pass the verification in Formula (\ref{eq:verify1}), since in $\pi_D$, $z_k^1=\sum x_iw_{ik}^1$. However, if $E_1'\neq E_1$, it can be detected due to the fact that $E_1'$ can only escape     the verification in Formula (\ref{eq:verify2}) with negligible probability. If $E_1'=E_1$ and $E_2'\neq E_2$, after the server updates the weight parameters and calculates the output, he is able to spot the erroneous $E_2'$ because $E_2'$ cannot pass the verification in Formula (\ref{eq:verify2}). 

\noindent{\em Case 4: $\pi_E'\neq \pi_E, \pi_D'\neq \pi_D, \pi_W'\neq \pi_W$}. The adversary $\mathbf{Adv}$ forges the proof $\pi_D'$, aiming to let $\pi_E'$ pass the verification. The proof $\pi_D'$ and $\pi_W'$ must satisfy two conditions: (1) $E_1'$ and $E_2'$ meet the convergence condition; and (2) $E_1'$, $E_2'$ and $\pi_D'$ and $\pi_W'$ pass the verification in Formula (\ref{eq:verify1} - \ref{eq:verify3}). 
Recall that $\pi_D'= \{\{g^{x_i'}\}, g^{y'}|\forall (\vec{x}, y)\in D\}$, and $\pi_W'=\{\{\Delta w_{jk}^{1'}\}, \{z_k^{1'}\}, g^{\delta^{o'}}, \{\delta_k^{L'}\}\}$. There are five possible approaches that $\mathbf{Adv}$ can launch to forge $\pi_D'$.

\hspace{0.1in}{\em Case 4.1}: $z_k^{1'}\neq z_k^1$ or $\hat{z}_k^{1'}\neq \hat{z}_k^1$, for some $1\leq k\leq d_1$. In this case, $\mathbf{Adv}$ provides incorrect linear activation on the first hidden layer of the DNN. With access to $\{g^{x_i}\}$, the server is capable of identifying this cheating behavior because $z_k^{1'}$ and $\hat{z}_k^{1'}$ cannot pass the verification by Formula (\ref{eq:verify1}).

\hspace{0.1in}{\em Case 4.2}: $g^{\delta^{o'}} \neq g^{\delta^o}$. Any incorrect error signal on the output layer must be detected. This is because the server learns $o$ and $z^o$ in the first feedforward procedure. As a consequence, according to the bilinearity property of bilinear pairing (Section \ref{sc:mapping}), the incorrect error signal must fail the verification by Formula (\ref{eq:verify4}). 

\hspace{0.1in}{\em Case 4.3}: $\delta_k^{L'} \neq \delta_k^L$ for some $1\leq k\leq d_L$. During the first feedforward computation, the server computes $z_k^{L-1}$. Therefore, if $\delta_k^{L'} \neq \delta_k^L$, it must fail the verification by Formula (\ref{eq:verify5}). 

\hspace{0.1in}{\em Case 4.4}: $\Delta w_{jk}^{1'}\neq \Delta w_{jk}^1$ for some $1\leq j\leq m$ and $1\leq k\leq d_1$. Since the server computes the correct error signal $\{\delta_k^1\}$ and obtains $\{g^{x_j}\}$ from $\pi_D'$, the incorrect weight increment  will fail the verification by Formula (\ref{eq:verify3}).

\hspace{0.1in}{\em Case 4.5}: $g^{x'_i}\neq g^{x_i}$ for some $i$ or $g^{y'}\neq g^{y}$. Because $\mathbf{Adv}$ is not allowed to manipulate $x_i$ or $y$, he can only cheat on the generator $g$. In particular, $\mathbf{Adv}$ uses a different generator $h$ to produce $h^{x_i}$ and $h^{y}$. Obviously, $h=g^{q}$, where $gcd(p,q)=1$, and $p$ is the order of the cyclic group $G$. Given $\Pi e(h^{x_i}, g^{w_{ik}^1})=e(g^{qx_i}, g^{w_{ik}^1})=e(g,g)^{z_k^1q}$, in order to pass the verification in Formula (\ref{eq:verify1}), $\mathbf{Adv}$ must set $z_k^{1'}=z_k^1q$. However, manipulating the linear activation of the first hidden layer completely changes the behavior of the neural network, including the number of epochs to reach convergence, the output and error at convergence. Therefore $\mathbf{Adv}$ cannot come up with $E_1'$, $E_2'$ and $\{\frac{\partial C}{\partial w_{jk}^1}'\}$ to pass the verification in Formula (\ref{eq:verify2}) and (\ref{eq:verify3}) without executing the training process. 
$\mathbf{Adv}$ can do the following. 
First, it generates $r, s\in Z_p$ s.t. $gcd(r, p)=1$.
Second, it creates the fake validation data $\vec{x}'=r\vec{x}$ and $y'=sy$. 
Third, it trains the DNN with the initial parameters $W$ and fake validation data $(\vec{x}', y')$.
It obtains two consecutive validation error $E_1'$ and $E_2'$ at convergence, as well as the weight update $\Delta L'$.
Finally, it constructs the proof $\pi_E'=\{g^{E_1'}, g^{E_2'}\}$, $\pi_D'=\{\{h_1^{x_i}\}, h_2^{y}| \forall (\vec{x}, y)\in V\}$ and $\pi_W=\{\{\Delta w_{jk}^{1'}\}, \{z_k^{1'}\}, \{\hat{z}_k^{1'}\}, g^{\delta^{o'}}, \{\delta_k^{L'}\}\}$, where $h_1=g^r$, $h_2=g^s$, $z_k^{1'}=\sum x_i' w_{ik}^1$, $\{\Delta w_{jk}^{1'}\}$, $\{\delta^{o'}_k\}$ and $\{\delta_k^{L'}\}$ are weight increment and the error signals at convergence with $\{\vec{x}', \vec{y}'\}$. 
Obviously, $\Delta L'$ and $\pi'$ can pass the verification process. 
However, we argue that this attack is equivalent to the process that $\mathbf{Adv}$  uses incorrect input data $\{\vec{x}', y'\}$. This violates our assumption that data integrity has been verified. 

\nop{
\subsection{Defend against possible inference attack}
\Bo{Since the bilinear mapping is deterministic, the same feature values will lead to same signatures, thus the dataset is vulnerable to inference attack. Particularly, the adversary can apply frequency-analysis attack to the signature and infer the original dataset values. To address this problem, we extend our framework and allow the participants to vary the generator in different rounds of model updates so that the same feature values can corresponds to different signatures, and the server follows the same verification protocol to verify the local model integrity. In this way, it prevents curious server to infer the data values by applying frequency-analysis attack with only minor communication cost between participants and server.}
}

\nop{
\noindent{\bf Proof Sketch.} Due to the space limit, we provide the proof sketch here. The formal proof can be found in the Appendix. Let $\Delta L'$ be the incorrect local model update provided by an adversary $\mathbf{Adv}$. 
If $\mathbf{Adv}$ provides the correct proof $\pi$, it must fail the verification in Formula (\ref{eq:verify1}), since $z_k^1$ must also be changed accordingly to pass the verification.
Otherwise, $\mathbf{Adv}$ generates the incorrect proof $\pi_E'=\{E_1', E_2'\}$, $\pi_W'$ and $\pi_D'$ such that $E_1'$ and $E_2'$ meet the convergence condition. This can be verified by $\pi_W'$ and $\pi_V'$. The only feasible solution is that $\mathbf{Adv}$ trains the local model by using wrong input data, which contradicts our assumption in Section \ref{sc:attack}. Therefore, we can conclude that \system enables the server to catch any incorrect local model update.\qed
}
\nop{
\begin{table}[!hbtp]
    \centering
    \begin{tabular}{|c|c|}
        \hline
        Basic Operation  & Verification  \\\hline
        Formula (\ref{eq:z}) (When $\ell=1$)  & Formula (\ref{eq:verify1}) \\\hline
        Formula (\ref{eq:cost} and \ref{eq:error}) & Formula (\ref{eq:verify2}) \\\hline
        Formula (\ref{eq:signal_1}) & Formula (\ref{eq:verify4}) \\\hline
        Formula (\ref{eq:signal_2}) (When $\ell=L$) & Formula (\ref{eq:verify5}) \\\hline
        Formula (\ref{eq:gradient}) (When $\ell=1$) & Formula (\ref{eq:verify3}) \\\hline
    \end{tabular}
    \caption{Relationship between basic DNN operations in Section \ref{sc:basic} with verification in Section \ref{sc:verify}}
    \label{tb:replace}
\end{table}
}

\nop{
\begin{proof}

Let $\Delta L$ be the correct local model update, and $\pi=\{\pi_E, \pi_D, \pi_W\}$ the correct proof of $\Delta L$.
Suppose an adversary $\mathcal{A}$ returns an incorrect local update $\Delta L'\neq \Delta L$ and the proof $\pi'=\{\pi_E', \pi_D', \pi_W'\}$. We assume that $\mathcal{A}$ is fully aware of the details of our authentication method. Thus it will try to generate $\pi'$ intelligently, trying to escape from the authentication. According to Formula (\ref{eq:verify3}), $\pi_W$ and $\pi_D$ must be consistent in order to pass the verification. There are three types of possible adversarial behaviors by $\mathcal{A}$: (1) $\pi_E'= \pi_E$, $\pi_D'=\pi_D$, and $\pi_W'=\pi_W$ (i.e., correct errors and correct proof); (2) $\pi_E'\neq \pi_E, \pi_D'=\pi_D, \pi_W'=\pi_W$ (i.e., incorrect errors and correct proof); and (3) $\pi_E'\neq \pi_E, \pi_D'\neq \pi_D, \pi_W'\neq \pi_W$ (i.e., incorrect errors and incorrect proofs). Next, we discuss these behaviors case by case, and show that for each case, the incorrect local update can escape from the authentication with zero probability (Def. \ref{def:security}).

\noindent{\bf Case 1: $\pi_E'= \pi_E$, $\pi_D'=\pi_D$, and $\pi_W'=\pi_W$}. For this case, the verification by Formula (\ref{eq:verify1}) must fail, since $z_k^1$ is generated by using the weights with correct update. It cannot match with $z_k^1$ by using incorrect weights. 

\noindent{\bf Case 2: $\pi_E'\neq \pi_E, \pi_D'=\pi_D$, and $\pi_W'=\pi_W$}. The adversary $\mathcal{A}$ returns incorrect $\pi_E'=\{E_1', E_2'\}$, where either $E_1'\neq E_1$ or $E_2'\neq E_2$, and the correct proof $\pi_D$ constructed from the right $E_1$ and $E_2$. Then $\mathcal{A}$ can pass the verification in Formula (\ref{eq:verify1}), since in $\pi_D$, $z_k^1=\sum_{x_iw_{ik}^1}$. However, if $E_1'\neq E_1$, it can be detected with 100\% probability due to the fact that $E_1'$ must fail the verification in Formula (\ref{eq:verify2}). If $E_1'=E_1$ and $E_2'\neq E_2$, after the server updates the weight parameters and calculates the output, he is able to spot the erroneous $E_2'$ because $E_2'$ cannot pass the verification in Formula (\ref{eq:verify2}). 

\noindent{\bf Case 3: $\pi_E'\neq \pi_E, \pi_D'\neq \pi_D, \pi_W'\neq \pi_W$}. The adversary $\mathcal{A}$ forges the proof $\pi_D'$, aiming to let $\pi_E'$ pass the verification. The proof $\pi_D'$ and $\pi_W'$ must satisfy two conditions: (1) $E_1'$ and $E_2'$ meet the convergence condition; and (2) $E_1'$, $E_2'$ and $\pi_D'$ and $\pi_W'$ pass the verification in Formula (\ref{eq:verify1} - \ref{eq:verify3}). 
Recall that $\pi_D'= \{\{g^{x_i'}\}, g^{y'}, \{z_k^{1'}\}, g^{\delta^{o'}}, \{\delta_k^{L'}\}|\forall (\vec{x}, y)\in D\}$, and $\pi_W'=\{\Delta w_{jk}^{1'}\}$. There are four possible approaches that $\mathcal{A}$ can launch to forge $\pi_V'$.

\hspace{0.2in}{\bf Case 3.1}: $z_k^{1'}\neq z_k^1$ or $\hat{z}_k^{1'}\neq \hat{z}_k^1$, for some $1\leq k\leq d_1$. In this case, $\mathcal{A}$ provides incorrect linear activation on the first hidden layer of the DNN. With access to $\{g^{x_i}\}$, the server is capable of identifying this cheating behavior because $z_k^{1'}$ and $\hat{z}_k^{1'}$ cannot pass the verification by Formula (\ref{eq:verify1}).

\hspace{0.2in}{\bf Case 3.2}: $g^{\delta^{o'}} \neq g^{\delta^o}$. Any incorrect error signal on the output layer must be detected. This is because the server learns $o$ and $z^o$ in the first feedforward procedure. As a consequence, according to the bilinearity property of bilinear pairing (Section \ref{sc:mapping}), the incorrect error signal must fail the verification by Formula (\ref{eq:verify4}). 

\hspace{0.2in}{\bf Case 3.3}: $\delta_k^{L'} \neq \delta_k^L$ for some $1\leq k\leq d_L$. Similar to the analysis in Case 3.3, during the first feedforward computation, the server computes $(z_k^{L-1}$. Therefore, if $\delta_k^{L'} \neq \delta_k^L$, it must fail the verification by Formula (\ref{eq:verify5}). 

\hspace{0.2in}{\bf Case 3.4}: $\Delta w_{jk}^{1'}\neq \Delta w_{jk}^1$ for some $1\leq j\leq m$ and $1\leq k\leq d_1$. Since the server computes the correct error signal $\{\delta_k^1\}$ and obtains $\{g^{x_j}\}$ from $\pi_D'$, the incorrect weight increment  will fail the verification by Formula (\ref{eq:verify3}).

\hspace{0.2in}{\em Case 3.5}: $g^{x'_i}\neq g^{x_i}$ for some $i$ or $g^{y'}\neq g^{y}$. Because $\mathcal{A}$ is not allowed to manipulate $x_i$ or $y$, he can only cheat on the generator $g$. In particular, $\mathcal{A}$ uses a different generator $h$ to produce $h^{x_i}$ and $h^{y}$. Obviously, $h=g^{q}$, where $gcd(p,q)=1$, and $p$ is the order of the cyclic group $G$. Given $\Pi e(h^{x_i}, g^{w_{ik}^1})=e(g^{qx_i}, g^{w_{ik}^1})=e(g,g)^{z_k^1q}$, in order to pass the verification in Formula (\ref{eq:verify1}), $\mathcal{A}$ must set $z_k^{1'}=z_k^1q$. However, manipulating the linear activation of the first hidden layer completely changes the behavior of the neural network, including the number of epochs to reach convergence, the output and error at convergence. Therefore $\mathcal{A}$ cannot come up with $E_1'$, $E_2'$ and $\{\frac{\partial C}{\partial w_{jk}^1}'\}$ to pass the verification in Formula (\ref{eq:verify2}) and (\ref{eq:verify3}) without executing the training process. 
For example, $\mathcal{A}$ can do following. 
First, it generates $r, s\in Z_p$ s.t. $gcd(r, p)=1$. 
Second, it creates the fake validation data $\vec{x}'=r\vec{x}$ and $y'=sy$. 
Third, it trains the DNN with the initial weight parameters $W$ and fake validation data $(\vec{x}', y')$.
It obtains two consecutive validation error $E_1'$ and $E_2'$ at convergence, as well as the weight update $\Delta L'$.
Finally, it constructs the proof $\pi_E'=\{g^{E_1'}, g^{E_2'}\}$, and $\pi_D'=\{\{h_1^{x_i}\}, h_2^{y}, \{z_k^{1'}\}, \{\hat{z}_k^{1'}\}, g^{\delta^{o'}}, \{\delta_k^{L'}\}| \forall (\vec{x}, y)\in V\}$ and $\pi_W=\{\Delta w_{jk}^{1'}\}$, where $h_1=g^r$, $h_2=g^s$, $z_k^{1'}=\sum x_i' w_{ik}^1$, $\{\Delta w_{jk}^{1'}\}$, $\{\delta^{o'}_k\}$ and $\{\delta_k^{L'}\}$ are weight increment and the error signals at convergence with $\{\vec{x}', \vec{y}'\}$. 
Obviously, $\Delta L'$ and $\pi'$ can pass the verification process. 
However, we argue that this attack is equivalent to the process that $\mathcal{A}$  uses incorrect input data $\{\vec{x}', y'\}$. This violates our assumption that the participant cannot cheat on data integrity. Thus, such attack can be caught by data integrity verification. 

With all the cases discussed, we conclude that \system is secure.
\end{proof}
}

\vspace{-0.1in}
\subsection{Proof of Lemma 4.1}
\noindent{\bf Proof.} For the sake of simplicity, we assume that $u_1$ is the only negative value in $\vec{u}$ and $\vec{v}$. It is easy to see that 

\vspace{-0.2in}
\begin{equation*}
\begin{split}
    \sum [u_i][v_i] &= [u_1][v_1] + [u_2][v_2] + \dots + [u_m][v_m] \\
    & = (u_1+p)v_1 + u_2v_2 + \dots + u_m v_m \\
    & = z + pv_1. 
\end{split}
\vspace{-0.05in}
\end{equation*}
If $z\geq 0$, $[z+pv_1] = z$; otherwise, since $-p<z<p$, $[z+pv_1] = z + p$. 
}

\end{document}